\documentclass[prd,onecolumn,notitlepage,showpacs,preprintnumbers,amsmath,amssymb,nofootinbib,aps,10pt,superscriptaddress]{revtex4-2}

\usepackage{dcolumn}
\usepackage{bm}
\usepackage[dvipsnames]{xcolor}
\usepackage[spanish,english]{babel}
\usepackage[utf8]{inputenc}
\usepackage{amsmath,amssymb,amsfonts,latexsym,cancel}
\usepackage{graphicx}
\usepackage{color}
\usepackage{soul}
\usepackage[normalem]{ulem}
\usepackage{slashed}
\usepackage{comment}
\usepackage{subfig}
\usepackage{float} 
\usepackage{mathtools}
\usepackage{mathrsfs}
\usepackage{braket}
\usepackage{enumitem}
\usepackage[many]{tcolorbox}

\usepackage{newtxtext}
\usepackage[varvw]{newtxmath}
\usepackage{array}
\usepackage{etoolbox}
\usepackage{ragged2e}
\usepackage[colorlinks=true,linkcolor=blue,citecolor=blue,urlcolor=cyan,plainpages=false,pdfpagelabels]{hyperref}

\newcommand{\eqtext}[1]{\ensuremath{\stackrel{\text{#1}}{=}}}

\newcommand{\p}{\partial}

\newcommand{\om}{\omega}

\newcommand{\pr}{\prime}
\newcommand{\testcmd}[2]{{}^#1\hspace{-1.6mm}{}_#2}
\DeclareMathOperator{\sgn}{sgn}

\DeclareMathAlphabet\mathbfcal{OMS}{cmsy}{b}{n}

\begin{document}

\title{{\bf An introduction to nonlinear fiber optics and optical analogues to gravitational phenomena}}

\author{Dimitrios Kranas}\email{dimitrioskranas@gmail.com}

\affiliation{Universidad Carlos III de Madrid, Departamento de Matem\'{a}ticas.
Avenida de la Universidad 30 (edificio Sabatini), 28911 Legan\'{e}s (Madrid), Spain}

\affiliation{Laboratoire de Physique de l’Ecole Normale Sup\'erieure, ENS, CNRS, Universit\'e PSL, Sorbonne Universit\'e, Universit\'e Paris Cit\'e, 75005 Paris, France}

\affiliation{Department of Physics and Astronomy, Louisiana State University, Baton Rouge, LA 70803, U.S.A.}

\author{Andleeb Zahra}\email{az65@st-andrews.ac.uk}

\affiliation{School of Physics and Astronomy, SUPA, University of St. Andrews,
North Haugh, St. Andrews, KY16 9SS, United Kingdom}

\author{Friedrich K\"{o}nig}\email{fewk@st-andrews.ac.uk}

\affiliation{School of Physics and Astronomy, SUPA, University of St. Andrews,
North Haugh, St. Andrews, KY16 9SS, United Kingdom}

\date{\today}

\begin{abstract}

The optical fiber is a revolutionary technology of the past century. It enables us to manipulate single modes in nonlinear interactions with precision at the quantum level without involved setups. This setting is useful in the field of analogue gravity, where gravitational phenomena are investigated in accessible analogue lab setups. These lecture notes provide an account of this analogue gravity framework and applications. Although light in nonlinear dielectrics is discussed in textbooks, the involved modelling often includes many assumptions that are directed at optical communications, some of which are rarely detailed. Here, we provide a self-contained and sufficiently detailed description of the propagation of light in fibers, with a minimal set of assumptions, which is relevant in the context of analogue gravity. Starting with the structure of a step-index fiber, we derive linear-optics propagating modes and show that the transverse electric field of the fundamental mode is well approximated as linearly polarized and of a Gaussian profile. We then incorporate a cubic nonlinearity and derive a general wave envelope propagation equation. With further simplifying assumptions, we arrive at the famous nonlinear Schr\"odinger equation, which governs fundamental effects in nonlinear fibers, such as solitons. As a first application in analogue gravity, we show how intense light in the medium creates an effective background spacetime for probe light akin to the propagation of a scalar field in a black hole spacetime.  We introduce optical horizons and particle production in this effective spacetime, giving rise to the optical Hawking effect. Furthermore, we discuss two related light emission mechanisms. Finally, we present a second optical analogue model for the oscillations of black holes, the quasinormal modes, which are important in the program of black hole spectroscopy \cite{berti2025}.

\end{abstract}

\maketitle


{
\hypersetup{linkcolor=red!40!black}
\tableofcontents
}


\section{Introduction}
\label{sec:intro}

A series of fascinating theoretical predictions involving black hole physics and cosmology, such as the Hawking effect, suffer from the weak character of their signals, rendering them practically unobservable. For example, the Hawking temperature of a solar mass black hole is approximately $60$ nk. This is $8$ orders of magnitude lower than the cosmic microwave background temperature, indicating that Hawking radiation is obscured by the ambient cosmic signal. To this end, analogue gravity, the research field studying phenomena of field theory in gravitational physics in equivalent non-gravitational platforms (analogue models), offers a great alternative pathway to test many of these exciting predictions, as it is experimentally accessible. The generating mechanism enabling this correspondence are similarities between the propagation equation of field perturbations in dispersive media, with those in a curved spacetime background, as was originally demonstrated in \cite{Unruh1981}. Thus, the task of analogue gravity is to build on the correspondence between curved geometries and dispersive media to generalize phenomena, initially thought to be associated only with gravity. Their experimental verification in the lab thus provides a deeper insight into them \cite{Schutzhold2025}. 

The most well-studied analogue setups involve two main families: 1) acoustic black holes \cite{Visser1998, Jacobson, Num4, euve2020} such as flowing water in a tank \cite{water.Unruh, water.noise, water.negative.freq, coutant2016} or Bose-Einstein condensates (BEC) \cite{sonic.BH.BEC, Barcelo2001, BH.realization, BEC.entanglement.Parentani, Jeff.theory, Jeff.nature, munoz2019, eckel2018, posazhennikova2016}. 2) Optical black holes \cite{Leonhardt2008, filaments.experiment.PRL,  Num3, Unruh, Leonhardt2019}. Other analogue systems include polariton fluids in microcavities \cite{polariton, Num8, nguyen2015}, and even superconducting electric circuits \cite{superconducting.chips}. The interested reader is referred to \cite{TheoryReview} for a nice review of the theoretical progress of the field in the first 20 years, and an experimental review can be found in \cite{ExpReview}.

In this work, we review the optical setup in which light propagating in a dispersive medium, such as a fiber, interacts nonlinearly with the molecules of the medium, thus altering the index of refraction locally. Consequently, test fields propagating thereon experience a varying refractive index, which generates optical horizons, and similarly to the black hole horizons, these produce entangled Hawking pairs of photons. Among the main experimental successes, one can include the generation of optical horizons \cite{Leonhardt2008, elazar2012} and the observation of the negative frequency mode \cite{Leonhardt2019}, participating in the Hawking effect. As the family of analogue gravity enjoys an ever-increasing number of advances, there are important milestones awaiting to be reached, including the observation of entanglement, confirming thus, the quantum nature of the particle creation processes. 

Due to the weak character of these particle creation phenomena, even in analogue frameworks, the idea of stimulating the process has been proposed and already experimentally studied in optical platforms \cite{Leonhardt2008,Leonhardt2019}, as well as water tank experiments \cite{water.Unruh, water.noise}. As these processes have been regarded as a classical amplification \cite{Leonhardt2019, water.Unruh}, one should seek different input states to enhance the quantum signal. In this spirit, the use of single-mode squeezed states has been proposed \cite{ABK,BAK} to amplify the amount of entanglement of the optical Hawking effect, paving the path to a promising experimental protocol towards observation of quantum features of these particle creation processes. Optical systems are excellent candidates for this task as they offer a great advantage for the detection of quantum particles, due to the precise preparation and manipulation of quantum states of light and the absence of a thermal background\footnote{Thermal optical fluctuations of room temperature are too weak to excite the detectors. This holds even in the involved analogue gravity experimental settings \cite{robustness}.}. Beyond particle creation, the effective field theories in analogue gravity can model relativistic field theoretical effects in an experimentally accessible way. This may even reach beyond the perturbative regime and be hard to model.  

The purpose of this work is to provide detailed calculations leading to a nonlinear wave equation describing the propagation of light in fibers. We first revisit some foundational approximations and evidence their validity as rarely explained in the literature. We then derive variants of the nonlinear wave equation in fibers in a straightforward way with a minimal set of approximations. This consolidated treatment should clarify the relation between the variants of equations in the literature and help exemplify what they are capable of describing: we employ these wave descriptions to provide a pedagogical review of the optical analogue framework and to study, in a simplified formulation, two specific examples: the creation of the analogue of the event horizon in optics for the Hawking effect and the analogue of the ringdown of a black hole in optics via their fingerprint quasinormal mode emission. The paper is organized as follows: We first study thoroughly the propagation of light in optical fibers, deriving the dispersion relation and the fundamental, linearly polarized mode solution, demonstrating that its transverse component has a Gaussian profile. Next, we include the nonlinearity generated by a strong field, derive a set of propagation equations akin to the generalized nonlinear Schr\"odinger equation suitable for the different analogue gravity cases we explore in this review, and examine its solutions. In the following section, we discuss the similarities of the optical framework with the propagation in a black hole spacetime. In particular, we establish the analogy between the optical and event horizons, provide the quantization scheme, and discuss in a simplified manner, suitable for this pedagogical review, the scattering of optical modes involved in the optical Hawking effect. Then, we explore a second analogue model: the physics of quasinormal modes produced by black hole oscillations and how to create an analogue thereof in optics. We close this paper with a short discussion and concluding remarks.

\section{Linear propagation of light in fibers}
\label{sec:linear}

Let us consider a fiber, which is a concrete experimental configuration used in analogue gravity experiments \cite{Leonhardt2008, Leonhardt2019}. We are concerned with a step-index fiber, which is made of two regions: 1) the core with an index of refraction $n_1$ and 2) the cladding with an index $n_2<n_1$. The index of refraction is constant in each of the two regions, with a step transition between the core and the cladding.  

\begin{figure}[H]
\begin{center}
\includegraphics[scale=0.45]{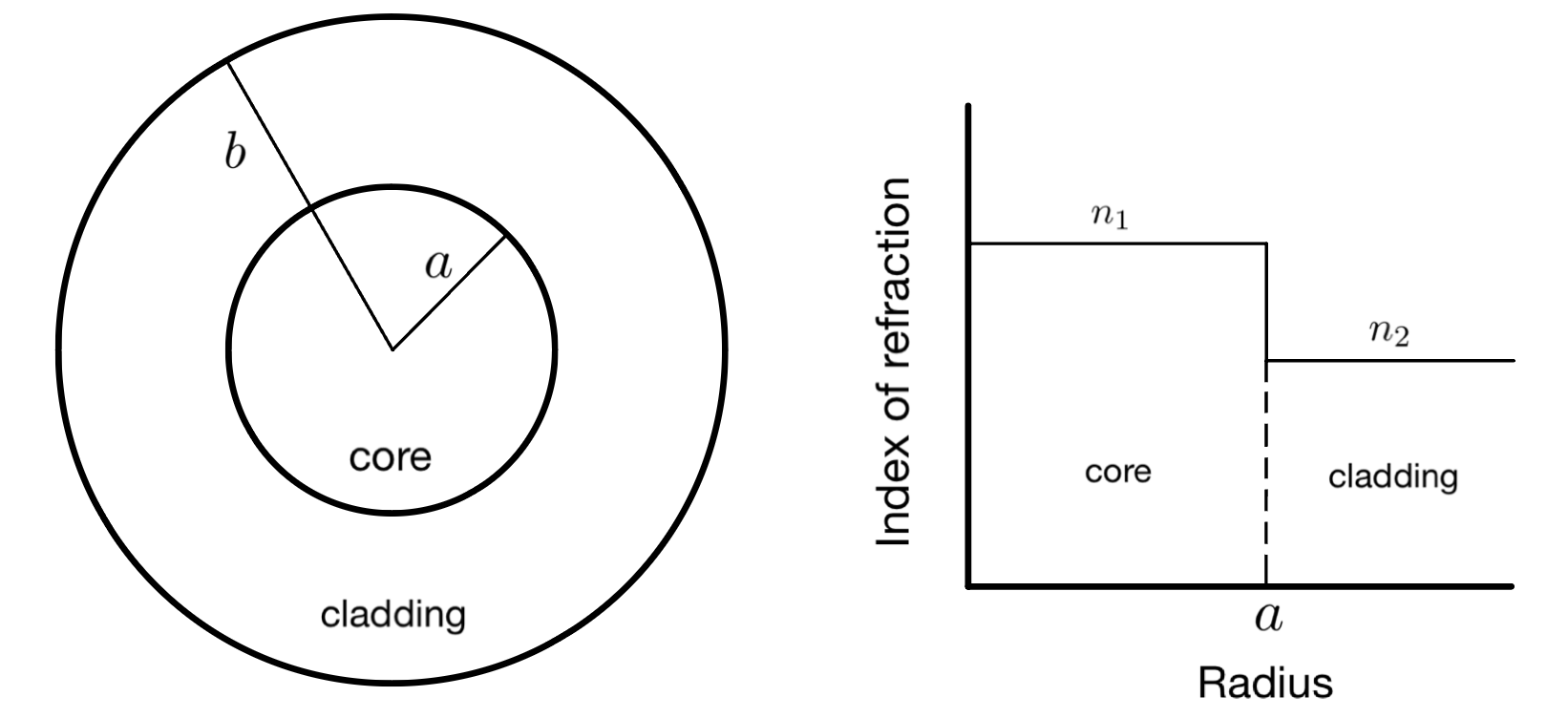}
\caption{Left panel: Cross-section of a step-index fiber, consisting of a core and a cladding surrounding it. Right panel: Index of refraction, modeled as a step-function with a value $n_1$ in the core and $n_2<n_1$ in the cladding. Typical value of the difference are $n_1-n_2\sim \mathcal{O}(10^{-3})$.}
\end{center}
\end{figure}  

\subsection{Derivation of the linear wave equation}

To study the available modes that can propagate along the fiber, we start with the Maxwell equations and the constitutive expressions. Then, to obtain the set of available modes, we apply Maxwell boundary conditions along the interface $\rho=a$ between the core and the cladding. Note that both regions are made of a dielectric medium and do not contain free charges or free currents, i.e. $\rho_\text{f}=0$, $\bm{J}_\text{f}=0$. In addition, none of the regions is magnetized and, thus, $\bm{M}=0$. Therefore, we have the following starting equations
\begin{align}
\nabla\cdot \bm{D}&=0, \label{M1}\\
\nabla\cdot \bm{B}&=0 \label{M2},\\
\nabla\times \bm{E}&=-\frac{\partial \bm{B}}{\partial{t}}, \label{M3} \\ 
\nabla\times \bm{H}&=\frac{\partial \bm{D}}{\partial t}, \label{M4}
\end{align}
where
\begin{align}
&\bm{D}=\epsilon_o \bm{E}+\bm{P}, \label{const1}\\
&\bm{B}=\mu_o \bm{H}, \label{const2}
\end{align}
where $\bm{E}$ is the electric field, $\bm{D}$ is the electric displacement, $\bm{P}$ is the material polarization vector, $\bm{B}$ is the magnetic field, $\bm{H}$ is the auxiliary magnetic field, $\epsilon_o$ is the vacuum permittivity and $\mu_o$ is the vacuum permeability. Taking the curl of (\ref{M3})
\begin{align}
\nonumber
\nabla\times\nabla\times \bm{E}&=-\frac{\partial}{\partial t}(\nabla\times \bm{B})\\
\nonumber
&\eqtext{(\ref{const2})} -\frac{\partial}{\partial t}(\mu_o \nabla\times \bm{H})\\
\nonumber
&\eqtext{(\ref{M4})}-\mu_o\frac{\partial}{\partial t}\left(\frac{\partial\bm{D}}{\partial t}\right)=-\mu_o\frac{\partial^2\bm{D}}{\partial t^2}\\
&\eqtext{(\ref{const1})}-\epsilon_o \mu_o \frac{\partial^2\bm{E}}{\partial t^2}-\mu_o\frac{\partial^2\bm{P}}{\partial t^2}.
\end{align}
Noting that the speed of light in vacuum is $c=\frac{1}{\sqrt{\epsilon_o \mu_o}}$, one obtains
\begin{align}
\nabla\times\nabla\times \bm{E}+ \frac{1}{c^2}\frac{\partial^2 \bm{E}}{\partial t^2}=-\mu_o\frac{\partial^2 \bm{P}}{\partial t^2} \label{wave1}
\end{align}

This is the most general expression for the wave equation. In general, the material polarization vector $P_i$ that the applied electric field $\bm{E}(\bm{r},t)$ induces can be expressed in the electric dipole approximation as the following series
\begin{align}
P_i(\bm{r},t)=\sum_{n=1}^{\infty}P_i^{(n)}(\bm{r},t)
\label{eq:polarizationextension}
 \end{align} 
 where
\begin{align} 
P_i^{(n)}({\bm{r},t})=\epsilon_o\int_{-\infty}^{\infty} dt_1 \cdots\int_{-\infty}^{\infty}dt_n \; \sum_{j_1,j_2,...,j_n} \chi^{(n)}_{ij_1\cdots j_n}(t-t_1,\cdots, t-t_n)E_{j_1}(\bm{r},t_1)\cdots E_{j_n}(\bm{r},t_n) \label{eq:polar}, 
\end{align}
where $\chi^{(n)}_{ij_1\cdots j_n}(t-t_1,\cdots, t-t_n)$ is a tensor of rank $n+1$, which is a function of the retarded times. The indices of the electric field and the polarization refer to their vector components. Relation (\ref{eq:polarizationextension}) is a relation of "cause-effect", where $\chi^{(n)}$ is the medium response. Due to causality we require $\chi^{(n)}(t-t')=0$ for $t'>t$. Furthermore, the response is assumed to be local. The incident electric field interacts with the medium which in turn generates an electric field of the same frequency and, in some cases, it excites modes at new frequencies. In this section, we only consider the linear response, for simplicity, to find the linearly polarized modes in the fiber. In the next section, we include the nonlinear response as well. The linear polarization term is given by
\begin{align}
&P^{(1)}_i(\bm{r},t)=\epsilon_o\int_{-\infty}^{t}dt^\prime \sum_{j}\chi^{(1)}_{ij}(t-t^\prime)E_j(\bm{r},t^\prime) \label{polar1}
\end{align}
Furthermore, we assume that both the core and the cladding are isotropic media, which leads to $\chi^{(1)}_{ij}(t^\prime-t)=\chi^{(1)}(t^\prime-t)\delta_{ij}$. Hence, the polarization vector field can be written as
\begin{equation}
    \bm{P}(\bm{r},t)=\epsilon_o\int_{-\infty}^{t}dt^\prime\chi^{(1)}(t-t^\prime)\bm{E}(\bm{r},t^\prime). \label{polar2}
\end{equation}
To proceed, we move to the Fourier space. The electric and polarization fields can be expanded as 
\begin{align}
\bm{E}(\bm{r},t)&=\int_{-\infty}^{+\infty}\frac{d\omega}{2\pi}\,\tilde{\bm{E}}(\bm{r},\omega)e^{-i\omega t}, \label{Four.E}\\
\bm{P}(\bm{r},t)&=\int_{-\infty}^{+\infty}\frac{d\omega}{2\pi}\,\tilde{\bm{P}}(\bm{r},\omega)e^{-i\omega t} \label{Four.P}
\end{align}
Since the polarization field $\bm{P}(\bm{r},t)$ is the time convolution of the susceptibility $\chi^{(1)}(t-t^\prime)$ and the electric field $\bm{E}(\bm{r},t^\prime)$, the Fourier transform of the polarization, by the convolution theorem, is the product of the Fourier transform of the susceptibility times the Fourier transform of the electric field, i.e.
\begin{equation}
\tilde{\bm{P}}(\bm{r},\omega)=\epsilon_o\tilde{\chi}^{(1)}(\omega)\tilde{\bm{E}}(\bm{r},\omega). \label{Four.P2} 
\end{equation}
From the latter, the constitutive relations in Fourier space read
\begin{align}
\tilde{\bm{D}}&=\epsilon_o[1+\tilde{\chi}^{(1)}(\omega)]\tilde{\bm{E}}, \label{const1.om}\\
 \tilde{\bm{B}}&=\mu_o\tilde{\bm{H}}, \label{const2.om}
\end{align}
where it should be understood that all Fourier modes of the fields above are functions of the frequency $\omega$ and space $\bm{r}$. Maxwell's equations in the Fourier domain are
\begin{align}
\nabla\cdot \tilde{\bm{D}}&=0, \label{M1.om}\\
\nabla\cdot \tilde{\bm{B}}&=0 \label{M2.om},\\
\nabla\times \tilde{\bm{E}}&=i\mu_o \om \tilde{\bm{H}}, \label{M3.om} \\ 
\nabla\times \tilde{\bm{H}}&=-i\om\tilde{\bm{D}}, \label{M4.om}
\end{align}
From (\ref{const1.om}) and (\ref{M1.om})\footnote{The inhomogeneity of the medium is included later in the form of boundary conditions.}, 
\begin{equation}
  \epsilon_o(1+\tilde{\chi}^{(1)})\nabla\cdot \tilde{\bm{E}}=0\Rightarrow \nabla\cdot \tilde{\bm{E}}=0.   
\end{equation}
The latter implies,
\begin{equation}
 \nabla \times \nabla \times \tilde{\bm{E}}=\nabla(\nabla\cdot \tilde{\bm{E}})-\nabla^2\tilde{\bm{E}}=-\nabla^2\tilde{\bm{E}}. \label{curl.curl.E}   
\end{equation}
In the Fourier domain and using (\ref{curl.curl.E}), the wave equation (\ref{wave1}) becomes
\begin{equation}
 -\nabla^2 \tilde{\bm{E}}-\frac{\omega^2}{c^2}\tilde{\bm{E}}=\frac{\omega^2}{c^2}(1+\tilde{\chi}^{(1)})\tilde{\bm{E}}   
\end{equation}
and we obtain the Helmholtz equation for the electric field
\begin{equation}
\nabla^2\tilde{\bm{E}}+\frac{\omega^2}{c^2}n^2\tilde{\bm{E}}=0, \label{Helm.E}   
\end{equation}
where  
\begin{equation}
n^2(\om)\equiv 1+\tilde{\chi}^{(1)}(\om) \label{eq:n}
\end{equation}
defines the chromatic index of refraction. Taking the curl of (\ref{M4.om}), we arrive at the Helmholtz equation for the magnetic field
\begin{equation}
 \nabla^2\tilde{\bm{H}}+\frac{\omega^2}{c^2}n^2(\omega)\tilde{\bm{H}}=0. \label{Helm.H}   
\end{equation}
The Helmholtz equations for the electric (\ref{Helm.E}) and magnetic (\ref{Helm.H}) fields define the two core equations for the subsequent analysis in this section. These will determine the spatial profile of the allowed modes and, consequently, the dispersion relation.

\subsection{Solutions to the wave equation in fibers}

Our plan is to solve equations (\ref{Helm.E}) and (\ref{Helm.H}) for the core and the cladding regions and then apply boundary conditions at the interface. For the core region $\rho<a$, equations (\ref{Helm.E}) and (\ref{Helm.H}) hold for $n=n_1$, while, for the cladding region, we need to take $n=n_2$. Due to the fiber symmetry, we will be using cylindrical coordinates $(\rho,\phi,z)$ with $x=\rho \cos\phi$ and $y=\rho \sin\phi$. 

Let us now describe our strategy to obtain the components of the electric and magnetic fields in the core and the cladding regions. Following the standard textbook treatment, we begin by solving the Helmholtz equations for the components of the fields along the direction of propagation, which we take to be along the $z$-axis. Once $\tilde{E}_{z}$ and $\tilde{H}_{z}$ are obtained from equations (\ref{Helm.E}) and (\ref{Helm.H}), the rest of the components are \textit{uniquely} determined by Maxwell equations (see below). Now, let us write equation (\ref{Helm.E}) in cylindrical coordinates for $\tilde{E}_{z}$, i.e.
\begin{equation}
 \frac{\p^2\tilde{E}_{z}}{\p\rho^2}+\frac{1}{\rho}\frac{\p \tilde{E}_{z}}{\p \rho}+\frac{1}{\rho^2}\frac{\p^2\tilde{E}_{z}}{\p \phi^2}+\frac{\om^2}{c^2}n_1^2\tilde{E}_{z}+\frac{\p^2 \tilde{E}_{z}}{\p z^2}=0.  \label{Helm.E.PDE}
\end{equation}
We proceed by separation of variables, i.e. $\tilde{E}_{z}=\tilde{E}_{o1}F_1(z)F_2(\rho)F_3({\phi})$, where $\tilde{E}_{o1}(\omega)$ is the amplitude of the electric field, and $F_1(z)$, $F_2(\rho)$, $F_3(\phi)$ determine the spatial profile of the mode. Then, 
\begin{align}
\underbrace{\frac{1}{F_2}\frac{d^2F_2}{d \rho^2}+\frac{1}{\rho F_2}\frac{d F_2}{d \rho}+\frac{1}{\rho^2 F_3}\frac{d^2F_3}{d \phi^2}+\frac{\omega^2}{c^2}n_1^2(\omega)}_{\beta^2}+\underbrace{\frac{1}{F_1}\frac{d^2F_1}{d z^2}}_{-\beta^2}=0, \label{eq:wave_separation}
\end{align}
where $\beta$ is just the separation constant, which will turn out to be the wavenumber along the direction of propagation, as we will soon see. The first underbrace of the previous equation is further written as
\begin{align}
\underbrace{\frac{\rho^2}{F_2}\frac{d^2F_2}{d \rho^2}+\frac{\rho}{F_2}\frac{d F_2}{d \rho}+\rho^2\frac{\omega^2}{c^2}n_1^2(\omega)-\rho^2\beta^2}_{m^2}+\underbrace{\frac{1}{F_3}\frac{d^2F_3}{d \phi^2}}_{-m^2}=0,
\end{align}
where $m$ is another separation constant. In summary, we arrive at the following three ODEs 
\begin{align}
 &\frac{d^2F_1}{dz^2}+\beta^2F_1=0, \label{ODE1}\\
 &\frac{d^2F_2}{d\rho^2}+\frac{1}{\rho}\frac{dF_2}{d\rho}+\left(\frac{\om^2}{c^2}n_1^2-\beta^2-\frac{m^2}{\rho^2}\right)F_2=0, \label{ODE2}\\
 &\frac{d^2 F_3}{d\phi^2}+m^2F_3=0. \label{ODE3}
\end{align}
The solution to these equations is 
\begin{align}
 F_1(z)&=Ae^{i\beta z}+Be^{-i\beta z}, \\
 F_2(\rho)&=C J_m(p_1\rho)+D N_m(p_1\rho),\\
 F_3(\phi)&=E\cos[m(\phi-\phi_o)],
\end{align}
where $p_1$ is given by
\begin{equation}
 p_1=\sqrt{\frac{\om^2}{c^2}n_1^2-\beta^2}. \label{p1}
\end{equation}
Choosing modes that propagate in the $+z$-direction, implying that we do not consider solutions of the form $e^{-i\beta z}$, imposes that $B=0$. Furthermore, we choose a reference frame such that $\phi_o=0$. Also, requiring $F_3(\phi+2\pi)=F_3(\phi)$, m has to be integer. Finally, we require that the solution is bounded in $0\leq\rho\leq a$. Since, $\lim_{\rho\rightarrow 0}N_{m}(p_1\rho)=+\infty$, we require $D=0$. Absorbing the constants $A$, $C$, and $E$ in $E_{o1}$, the $z$-component of the electric field is
\begin{equation}
\tilde{E}_{z}=\tilde{E}_{o1}J_m(p_1\rho)\cos(m\phi)e^{i\beta z}. \label{Ex1}    
\end{equation}
Similarly, the magnetic field is given by
\begin{equation}
\tilde{H}_{z}=\tilde{H}_{o1}J_m(p_1\rho)\sin(m\phi)e^{i\beta z}.
\end{equation}
The angular dependence of the magnetic field has to be of the form $\sin(m\phi)$ when the electric field has angular dependence given by $\cos(m\phi)$ due to boundary conditions (see discussion below). 

In the cladding region $a\leq\rho\leq b$, the solution to the radial part has the general form
\begin{equation}
 F_2(\rho)=AI_m(p_2\rho)+BK_m(p_2\rho),   
\end{equation}
where 
\begin{equation}
 p_2=\sqrt{\beta^2-\frac{\om^2}{c^2}n_2^2},   
\end{equation}
and $I_m$ is the first kind of the modified Bessel functions and $K_m$ is the second kind of the modified Bessel functions. We wish the EM field to decay radially. Since, $\lim_{\rho\rightarrow \infty}I_m(p_2\rho)=+\infty$, we require $A=0$, and, thus, the radial part in the cladding is given exclusively by the function $K_m$. The $z$-components of the electric and magnetic fields are given by
\begin{align}
\tilde{E}_{z}&=\tilde{E}_{o2}K_m(p_2\rho)\cos(m\phi)e^{i\beta z}, \label{Ex2}    \\
\tilde{H}_{z}&=\tilde{H}_{o2}K_m(p_2\rho)\sin(m\phi)e^{i\beta z}. \label{Hx2}
\end{align}

In summary, we solved the Helmholtz equations (\ref{Helm.E}) and (\ref{Helm.H}) for the $z$-components of the electric and magnetic fields, respectively, both in the core ($\rho\leq a$) and in the cladding ($a\leq \rho \leq b$). The resulting expressions are
\begin{align}
\tilde{E}_{z}=
\begin{cases}
\tilde{E}_{o1}\,J_m(p_1\rho)\,\cos{(m\phi)}\,e^{i\beta z}, \;\,0\leq\rho\leq a,\\
E_{o2}\,K_m(p_2\rho)\,\cos{(m\phi)}\,e^{i\beta z}, \,a\leq\rho\leq b,
\end{cases} \label{Ex}\\
\tilde{H}_{z}=
\begin{cases}
\tilde{H}_{o1}\,J_m(p_1\rho)\,\sin{(m\phi)}\,e^{i\beta z}, \;\,0\leq\rho\leq a,\\
H_{o2}\,K_m(p_2\rho)\,\sin{(m\phi)}\,e^{i\beta z}, \,a\leq\rho\leq b.
\end{cases} \label{Hx}
\end{align}
The rest of the components are uniquely determined by Maxwell's equations according to (see, for example, section 2.2.2 of \cite{Agra-FOS}) 
\begin{align}
\nonumber
\tilde{E}_{\rho,j}&=(-1)^{j-1}\frac{i}{p_j^2}\left(\beta\frac{\partial \tilde{E}_{z}}{\partial \rho}+\mu_o\frac{\omega}{\rho}\frac{\partial \tilde{H}_{z}}{\partial \phi}\right)\\
&=(-1)^{j-1}\frac{i}{p_j^2}\left[\tilde{E}_{oj}p_j\beta f_{j,m}^\prime(p_j\rho)+\tilde{H}_{oj}m\mu_o\frac{\om}{\rho}f_{j,m}(p_j\rho)\right]e^{i\beta z} \cos{(m\phi)},\label{Erhoj}\\
\nonumber
\tilde{E}_{\phi,j}&=(-1)^{j-1}\frac{i}{p_j^2}\left(\frac{\beta}{\rho}\frac{\partial \tilde{E}_{z}}{\partial \phi}-\mu_o\omega\frac{\partial \tilde{H}_{z}}{\partial \rho}\right)\\
&=(-1)^{j-1}\frac{i}{p_j^2}\left[-\tilde{E}_{oj}m\frac{\beta}{\rho} f_{j,m}(p_j\rho)-\tilde{H}_{oj}p_j\mu_o\omega f_{j,m}^\prime(p_j\rho)\right]e^{i\beta z} \sin{(m\phi)},\label{Ephij} \\
\nonumber
\tilde{H}_{\rho,j}&=(-1)^{j-1}\frac{i}{p_j^2}\left(\beta\frac{\partial \tilde{H}_{z}}{\partial \rho}-\epsilon_o n_j^2\frac{\omega}{\rho}\frac{\partial \tilde{E}_{z}}{\partial \phi}\right)\\
&=(-1)^{j-1}\frac{i}{p_j^2}\left[\tilde{H}_{oj}p_j\beta f_{j,m}^\prime(p_j\rho)+\tilde{E}_{oj}m\epsilon_o n_j^2\frac{\om}{\rho}f_{j,m}(p_j\rho)\right]e^{i\beta z} \sin{(m\phi)},\label{Hrhoj} \\
\nonumber
\tilde{H}_{\phi,j}&=(-1)^{j-1}\frac{i}{p_j^2}\left(\frac{\beta}{\rho}\frac{\partial \tilde{H}_{z}}{\partial \phi}+\epsilon_o n_j^2\omega\frac{\partial \tilde{E}_{z}}{\partial \rho}\right)\\
&=(-1)^{j-1}\frac{i}{p_j^2}\left[\tilde{H}_{oj}m\frac{\beta}{\rho} f_{j,m}(p_j\rho)+\tilde{E}_{oj}p_j\epsilon_o n_j^2 \omega f_{j,m}^\prime(p_j\rho)\right]e^{i\beta z} \cos{(m\phi)}, \label{Hphij}
\end{align} 
where $j=1$ corresponds to the solutions in the core region $\rho<a$ while $j=2$ refers to the fields in the cladding region $a\leq\rho\leq b$. The overall minus sign between the core vs the cladding solutions comes from the definitions $p_2^2=\beta^2-n_2^2\omega^2/c^2$ as opposed to $p_1^2=n_1^2\omega^2/c^2-\beta^2$. Finally, note that the prime denotes differentiation with respect to the argument of the function, and we define

\begin{align}
f_{j,m}(x)=
\begin{cases}
J_m(x),\; j=1,\\
K_m(x),\; j=2.
\end{cases}
\end{align}

Next, we need to apply boundary conditions at the interface $\rho=a$. Those result in the continuity of the electric and magnetic field components. For example, from the continuity of the $\tilde{E}_{\rho}$ component, we realize that in equation (\ref{Erhoj}) and (\ref{Hphij}), the terms $\p \tilde{E}_{z}/\p \rho$ and $\p \tilde{H}_{z}/\p \phi$ must have the same angular dependence such that one can factor out the function of $\phi$. If that is not the case, then the boundary condition $\tilde{E}_{\rho,1}(\rho=a,\phi,z)=\tilde{E}_{\rho,2}(\rho=a,\phi,z)$  (where $\tilde{E}_{\rho,1}$ is the field in the core and $\tilde{E}_{\rho,2}$ is the field in the cladding) would be $\phi$-dependent, and therefore, not satisfied throughout the interface. Thus, when $\bm{E}\propto \cos(m\phi)$ then $\bm{H}\propto \sin(m\phi)$ and vice versa. The continuity of the $\tilde{E}_z$ and $\tilde{H}_z$ yield, respectively,  
\begin{align}
\tilde{E}_{z,1}(\rho=a,\phi,z)=\tilde{E}_{z,2}(\rho=a,\phi,z)\Rightarrow \tilde{E}_{o2}&=\tilde{E}_{o1}\frac{J_m(p_1a)}{K_m(p_2a)}, \label{BC.Ex}\\
\tilde{H}_{z,1}(\rho=a,\phi,z)=\tilde{H}_{z,2}(\rho=a,\phi,z)\Rightarrow \tilde{H}_{o2}&=\tilde{H}_{o1}\frac{J_m(p_1a)}{K_m(p_2a)}. \label{BC.Hx}
\end{align}
The continuity of $\tilde{E}_{\phi}$ implies
\begin{equation}
\tilde{E}_{o1}\frac{m\beta}{a}\left(\frac{1}{p_1^2}+\frac{1}{p_2^2}\right)=-\tilde{H}_{o1}\mu_o\om\left[\frac{1}{p_1}\frac{J_m^\prime(p_1a)}{J_m(p_1a)}+\frac{1}{p_2}\frac{K_m^\prime(p_2a)}{K_m(p_2a)}\right], \label{BC.Ephi}
\end{equation}
where we made use of equations (\ref{BC.Ex}) and (\ref{BC.Hx}). Similarly, the continuity of $\tilde{H}_{\phi}$ at the boundary results in
\begin{equation}
\tilde{H}_{o1}\frac{m\beta}{a}\left(\frac{1}{p_1^2}+\frac{1}{p_2^2}\right)=-\tilde{E}_{o1}\epsilon_o\om\left[\frac{n_1^2}{p_1}\frac{J_m^\prime(p_1a)}{J_m(p_1a)}+\frac{n_2^2}{p_2}\frac{K_m^\prime(p_2a)}{K_m(p_2a)}\right]. \label{BC.Hphi}   
\end{equation}
Solving equation (\ref{BC.Hphi}) for $\tilde{H}_{o1}$ and plugging back into (\ref{BC.Ephi}), we obtain the very important expression
\begin{align}
\left[\frac{1}{p_1}\frac{J_m^\prime(p_1a)}{J_m(p_1a)}+\frac{1}{p_2}\frac{K_m^\prime(p_2a)}{K_m(p_2a)}\right]\left[\frac{n_1^2}{p_1}\frac{J_m^\prime(p_1a)}{J_m(p_1a)}+\frac{n_2^2}{p_2}\frac{K_m^\prime(p_2a)}{K_m(p_2a)}\right]=\frac{c^2 m^2 \beta^2}{\omega^2 a^2}\left(\frac{1}{p_1^2}+\frac{1}{p_2^2}\right)^2
\label{dispersion}
\end{align}
This equation constitutes the dispersion relation, which determines the propagation constant $\beta$ as a function of the frequency $\omega$ and the mode index $m$. It also depends on the core radius $a$ and the refractive indices $n_1$ and $n_2$, which are determined based on the material.  Equation (\ref{dispersion}) can be simplified under the weakly guiding approximation \cite{Agra-NL, Fundamentals}, $n_1\approx n_2$ $(\beta\approx \om n_1/c)$, leading to the following dispersion relation
\begin{equation}
\left[\frac{1}{p_1}\frac{J_m^\prime(p_1a)}{J_m(p_1a)}+\frac{1}{p_2}\frac{K_m^\prime(p_2a)}{K_m(p_2a)}\right]=\pm \frac{m}{a}\left(\frac{1}{p_1^2}+\frac{1}{p_2^2}\right). \label{dispersion2}
\end{equation}

Recall that the parameters $p_1$ and $p_2$ contain $\beta$. Once the wavenumber $\beta$ has been obtained from equation (\ref{dispersion2}), one can compute all the components of the electric and magnetic fields. First, we have to determine the constant $\tilde{H}_{o1}$ in terms of $\tilde{E}_{o1}$. From equation (\ref{BC.Ephi}), we obtain\footnote{We could also use (\ref{BC.Hphi}) to determine $\tilde{H}_{o1}$ in terms of $\tilde{E}_{o1}$ as we did earlier to obtain (\ref{dispersion}).} 
\begin{equation}
 \tilde{H}_{o1}=-\tilde{E}_{o1}\frac{m\beta}{\mu_o \om a}\frac{\frac{1}{p_1^2}+\frac{1}{p_2^2}}{\frac{1}{p_1}\frac{J_m^\prime(p_1a)}{J_m(p_1a)}+\frac{1}{p_2}\frac{K_m^\prime(p_2a)}{K_m(p_2a)}} \label{Ho1.from.Eo1}
\end{equation}
Note that one has the freedom to choose the value of $\tilde{E}_{o1}$. The amplitudes $\tilde{E}_{o2}$ and $\tilde{H}_{o2}$ are determined from equations (\ref{BC.Ex}) and (\ref{BC.Hx}), respectively, using also (\ref{Ho1.from.Eo1}). Next, the components $\tilde{E}_{\rho}$ and $\tilde{E}_{\phi}$ are determined by equations (\ref{Erhoj}) and (\ref{Ephij}). Finally, one can recover the Cartesian components $\tilde{E}_{x}$ and $\tilde{E}_{y}$, as we show in the following subsection. 

In summary, we have derived the set of equations (\ref{Ex})-(\ref{Hphij}) that determines the spatial profile of the electric and magnetic fields both in the core and the cladding regions of the fiber. In addition, applying matching conditions at the interface, and under the weakly-guiding approximation, we derived the dispersion relation (\ref{dispersion2}), from which the wavenumber is determined.

\subsection{Linearly polarized solution and the fundamental mode}

In this subsection, we are interested in studying solutions to the Helmholtz equation (\ref{Helm.E}) and (\ref{Helm.H}) corresponding to linearly polarized modes of the electromagnetic field. Such solutions will provide the basis for the analysis of the nonlinear mode propagation in the following sections. The Cartesian components of the electric field on the transverse plane are obtained by
\begin{align}
\tilde{E}_{x}&=\cos{\phi}\tilde{E}_{\rho}-\sin{\phi}\tilde{E}_{\phi}, \label{Exx}\\
\tilde{E}_{y}&=\sin{\phi}\tilde{E}_{\rho}+\cos{\phi}\tilde{E}_{\phi}. \label{Eyy}
\end{align}

In the weakly guiding regime, fibers can support,  approximately, linearly polarized mode solutions (LP). Note that, in general, equation (\ref{dispersion}) may have more than one root. Let $\ell$ be a label indicating the root. The wavenumber, and, therefore, a propagating mode along the fiber, is characterized by $\om$, $m$, and $\ell$. The lowest-order of those linearly polarized solution is for $m=1$ and $\ell=1$. This is the fundamental mode of a fiber and is denoted as $HE_{11}$ or $LP_{01}$ \cite{Fundamentals}. To illustrate that this mode is linearly polarized and has a Gaussian profile, we provide some characteristic plots below using (\ref{Exx}), (\ref{Eyy}), (\ref{Erhoj}), and (\ref{Ephij}).  

\begin{figure}[H]
\begin{center}
\includegraphics[scale=0.60]{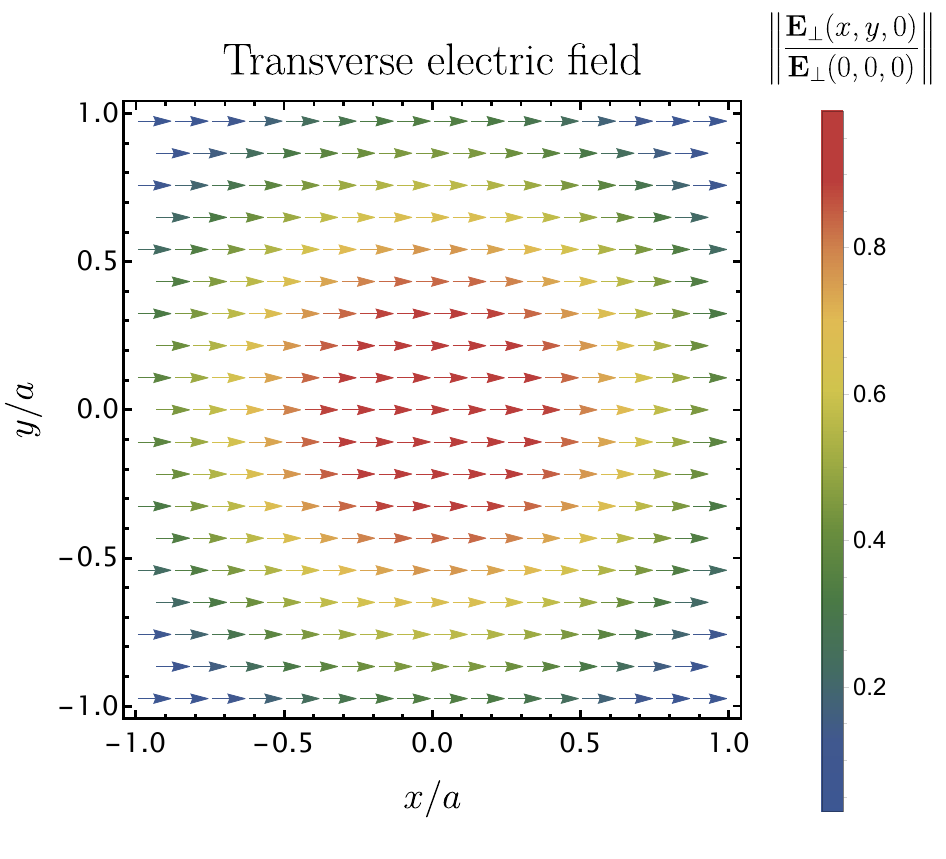}
\caption{Vector plot of the electric field, for $m=1$, $\ell=1$ (mode $HE_{11}$), on the transverse $x-y$ plane of a typical fiber. We use the following values: $a=5\text{$\,\mu$m}$, $\lambda_o=1\text{$\,\mu$m}$, (resulting in $\omega=1.884 \,\text{fs}^{-1}$), and fused silica as the medium for the cladding, resulting in $n_2=1.440$, and germanium doped fused silica for the core, resulting in $n_1=1.445$.}\label{fig.vector.plot}
\end{center}
\end{figure}  

It is evident from Fig.\ref{fig.vector.plot} that, to an excellent approximation,  the $HE_{11}$ mode is linearly polarized along the $x$-direction. Hence, the spatial profile of the electric field for that mode is of the form $\tilde{\bm{E}}(\bm{r})\approx \hat{x} \tilde{E}_x(\rho, \phi)$ where $\tilde{E}_x(\rho, \phi)$ is determined by equation (\ref{Exx}). Below, we plot the modulus of $\tilde{E}_x(\rho, \phi)$ on the transverse $x-y$ plane.

\begin{figure}[t]
\begin{center}
\includegraphics[scale=0.50]{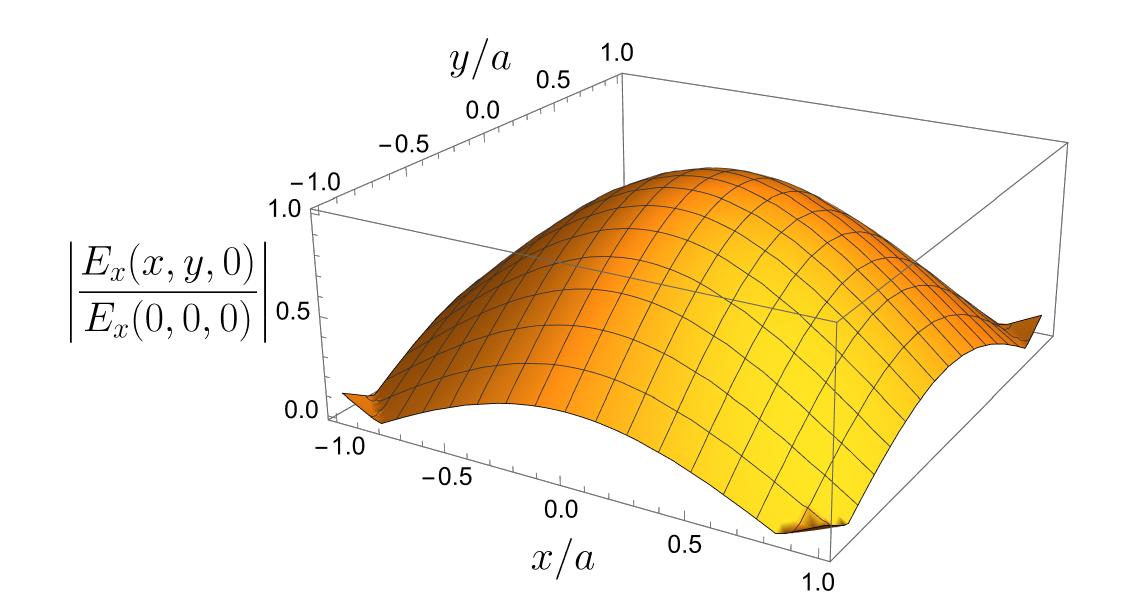}
\includegraphics[scale=0.365]{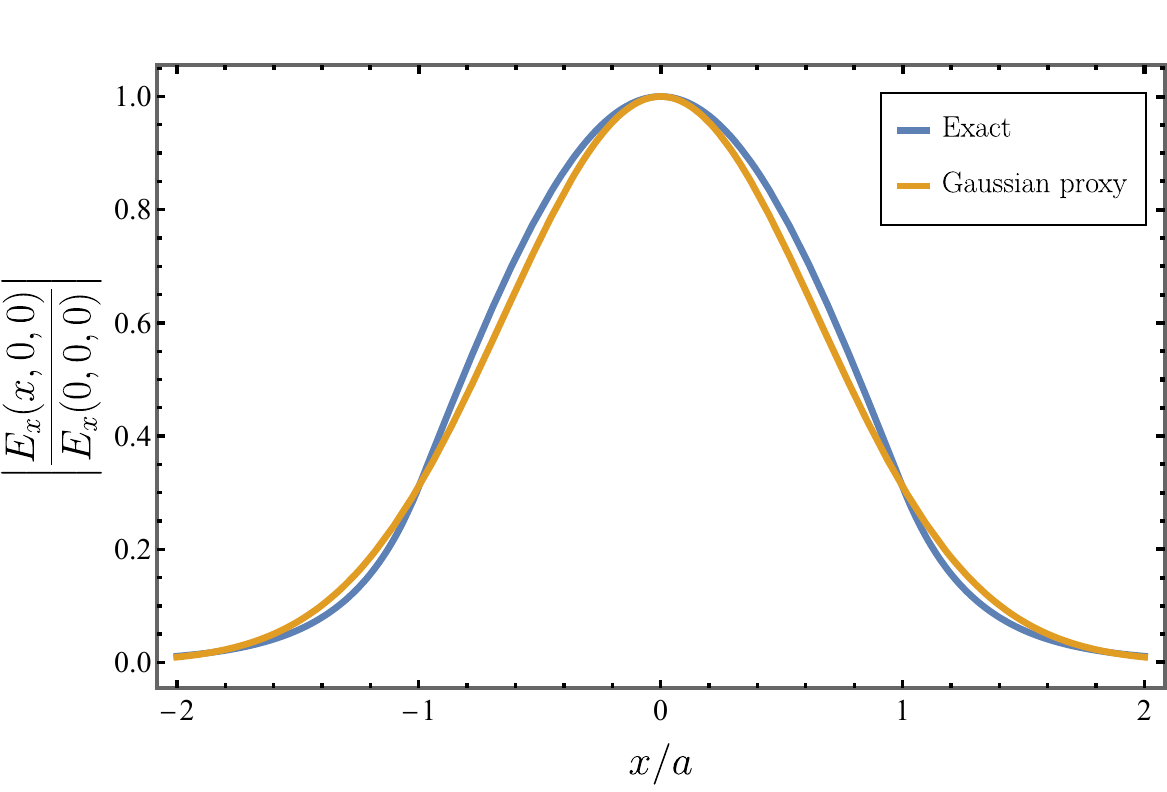}
\includegraphics[scale=0.35]{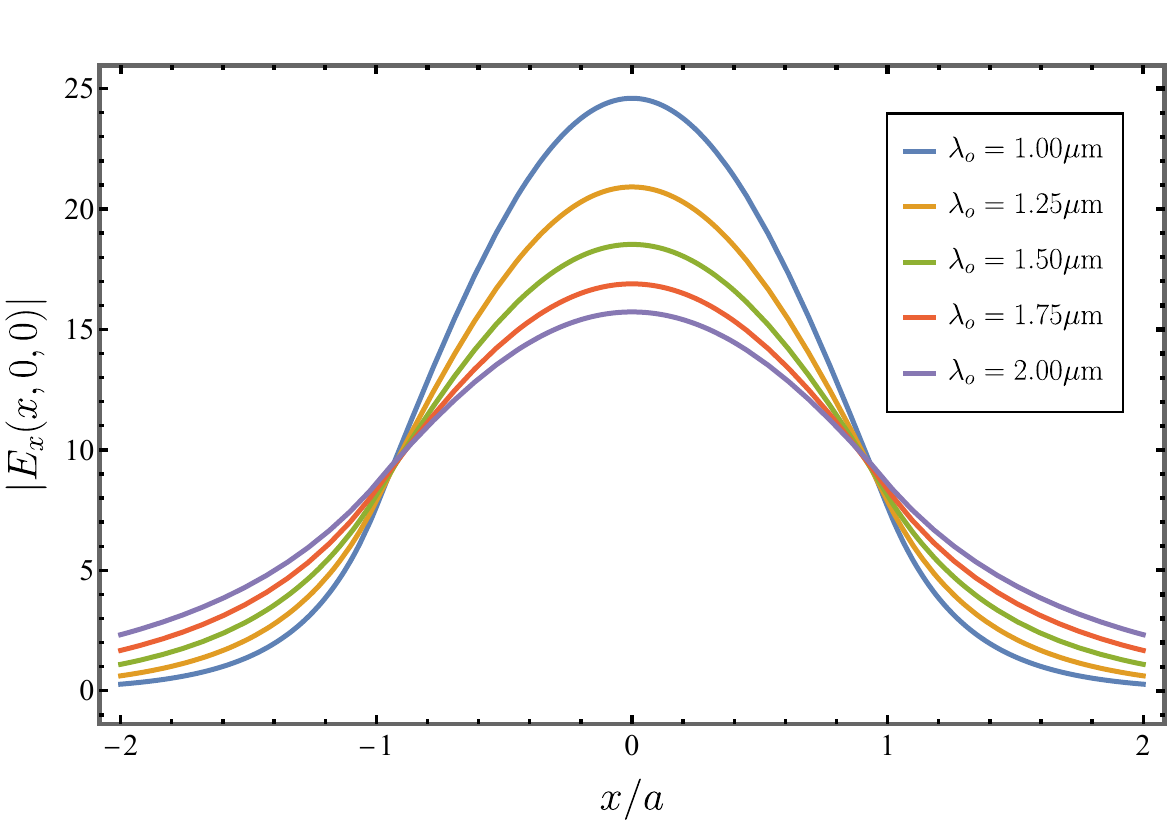}
\caption{\justifying $HE_{11}$ mode. Upper-left panel: 3D plot of the $E_x(x,y,0)$ component. Upper-right panel: Cross-section of the $E_x$ component with the $y=0$ and $z=0$ planes and comparison with the Gaussian function $|\tilde{E}_x(x,0,0)|/|\tilde{E}_x(0,0,0)|=\exp(-x^2/w^2)$. The best-fit parameter is $w=0.93 a$. Bottom panel: Cross-section of $E_x(x)$ with the $y=0$ and $z=0$ planes for different vacuum wavelengths. The vertical axis is expressed on the same units as the freely chosen amplitude $\tilde{E}_{o1}$. The Gaussian character is preserved as the frequency (or, equivalently, vacuum wavelength) varies, but with a different Gaussian peak and effective width. We use the same model parameters as in Fig.\ref{fig.vector.plot}. }\label{fig.Ex.transverse}
\end{center}
\end{figure}  

Fig.\ref{fig.Ex.transverse} shows that the mode $HE_{11}$ is axisymmetric, i.e. invariant under $\phi$-rotations, with a Gaussian-like fall-off in the $\rho$-direction. Because of that, the mode is very well-approximated by
\begin{equation}
\tilde{E}_x(x,y,z)=\tilde{E}_o e^{-\frac{x^2+y^2}{w^2}}e^{i\beta z}, \label{HE11}
\end{equation}
where $w$ determines the fall-off scale along the radial direction. In the following, we assume a Gaussian profile on the transverse plane for the fundamental mode.

We also illustrate the dependence of the mode profile on the wavelength of light. In Fig.\ref{fig.Ex.transverse} (bottom-right panel) we see that the distribution moderately widens when doubling the wavelength, i.e. the width remains approximately unchanged for small wavelength changes. We also observe that all profiles very nearly intersect at the boundary between core and cladding region, resulting in a constant intensity around $x=a$.

As a final remark, note that the number of modes supported by the fiber depends on its specifications, i.e. the radius of the core and the refractive indices $n_1$ and $n_2$. For given $\omega$ and $m$, the number of modes the fiber supports is the number of solutions to the dispersion relation (\ref{dispersion2}). For appropriate combinations of $a$, $n_1$, and $n_2$, the fiber supports a single mode: the fundamental mode $HE_{11}$. These fibers are referred to as \textit{single-mode fibers} and are the ones we consider in what follows. It is convenient to define the normalized frequency
\begin{equation}
V=a\frac{\om}{c}\sqrt{n_1^2-n_2^2}.    
\end{equation}
In terms of this parameter, the fiber supports only the fundamental mode when $V<2.405$ \cite{Agra-NL}. For $a=5\mu\text{m}$, the fiber supports only one mode for wavelengths satisfying $\lambda>1.5\mu\text{m}$. 

To summarize, this section provides the framework to study the propagation of light in fibers. In particular, we derived solutions to the Maxwell equations for the modes that the fiber supports and obtained the dispersion relation. Furthermore, we showed that the \textit{fundamental} mode, i.e. the first mode that the fiber can support, is, to a good approximation, linearly polarized and Gaussian. This approximation provides the basis of treating nonlinear effects in fibers, which we study in the following section.

\section{Nonlinear propagation of light in fibers}\label{sec:nonlinear_propagation}
We have derived a wave equation that governs the linear propagation of light in fibers, as well as the type of solutions supported therein. We now want to study the propagation of \textit{intense} light interacting with the medium in a nonlinear fashion. The nonlinear interaction is a weak effect in optical fibers and is treated perturbatively. The lowest non-vanishing nonlinear order of the susceptibility tensor in Eq. (\ref{eq:polarizationextension}) is $\chi^{(3)}$ as the $\chi^{(2)}$ tensor is zero due to the inversion symmetry of the fused silica glass \cite{Agra-NL, Boyd}. The $\chi^{(3)}$-term leads to a small local increase in the refractive index proportional to the intensity of light. This phenomenon is known as the \textit{optical Kerr effect} and it leads to a significant modification of the propagation equation, which we derive in this section. Furthermore, this refractive index change leads to the bending of light ray trajectories in a spacetime diagram, similar to the bending of rays due to gravity, i.e. this is the foundation of optical analogues.
Derivations of nonlinear wave equations can alternatively be found in the literature, e.g. \cite{Agra-NL, Boyd, conforti2013, bermudez2023, Leonhardt2008, finazzi2013, hasegawa1973}.

\subsection{Nonlinear propagation equations}
Our starting point is the propagation equation (\ref{wave1}), in which we now split the polarization into a linear and a nonlinear part, $\bm{P}=\bm{P}_{\rm L}+\bm{P}_{\rm NL}$. We also make use of relation (\ref{curl.curl.E}), which we justify in appendix \ref{sec:div_E}.
As in the previous section, we approximate the electric field and the optical polarization to be linear in the (transverse) x-direction, which leads to a scalar field description:
\begin{align}
\nabla^2 E(\bm{r},t)-\frac{1}{c^2}\frac{\partial^2 E(\bm{r},t)}{\partial t^2}=\mu_o\frac{\partial^2 P_{\rm L}(\bm{r},t)}{\partial t^2} + \mu_o\frac{\partial^2 P_{\rm NL}(\bm{r},t)}{\partial t^2}. \label{wave3}
\end{align}
The treatment can easily be generalized to describe the optical polarization by reintroducing the other field components. 
In section \ref{sec:linear}, we have seen that the transverse field profiles separate and that they are only weakly depending on $\omega$ (cf. Fig. \ref{fig.Ex.transverse}), i.e. on small changes in the refractive index, so that we will assume constant profiles.  
The optical Kerr effect also generates weak modifications of the refractive index, and thus we conclude that the separation is fulfilled to a good approximation, i.e.:
\begin{align}
E(\bm{r},t)=F(x,y) E(z,t) \quad
P_{\rm L}(\bm{r},t)=F(x,y) P_{\rm L}(z,t) \quad
P_{\rm NL}(\bm{r},t)\approx F(x,y) P_{\rm NL}(z,t)
, \label{eq:separation}
\end{align}
where we keep the same notation for the separated fields \footnote{For the nonlinear polarization this is an approximation as $P_{\rm NL}$ depends on the third power of $E$ (cf. Eqn. \ref{eq:nondelayedchi3}). Although the transverse profile of $P_{\rm NL}$ is the third power of the linear transverse profile, we approximate it as the same Gaussian, which is justified in view of the perturbative nature of the nonlinearity, which does not change the mode to first order \cite{Agra-NL}. }\footnote{$P_{\rm NL}(\bm{r},t)$ and $ P_{\rm NL}(z,t)$ are different functions but we name them the same for simplicity.}. 

For the linear description, the wave propagation is governed by $\chi^{(1)}$, the refractive index, or the propagation constant $\beta$. We now clarify their relation.
For the lossless medium, the susceptibility is connected to the refractive index by $\tilde{\chi}^{(1)}(\omega)$ $=
({n}(\omega)^2-1)$ (cf. equations (\ref{polar2}) and (\ref{eq:n})), leading to the following convolution:
\begin{align}
\frac{P_{\rm L}(z,t)}{\epsilon_0}= \int dt' \, \chi^{(1)}(t-t')\,E(z,t')=\int\,\frac{d \omega}{2 \pi}({n}(\omega)^2-1)\,\tilde{E}(z,\omega) \, e^{-i\omega t}. 
 \label{eq:susceptibilitytoindex}
\end{align}
We can also make the connection to the linear propagation constant $\beta$ of the fiber in Eq. (\ref{dispersion}) with the use of an effective refractive index defined as $\beta(\omega)={n}_{\mathrm{eff}}(\omega) \omega/c$. In the time domain this writes as: $\hat{\beta}(i\partial_t)=\hat{{n}}_{\mathrm{eff}}(i\partial_t)i\partial_t/c$, where $\partial_t$ is shorthand for $\partial/\partial_t$. Finally, we develop $\beta$ into a Taylor series around a frequency of interest:
\begin{align}
 \beta(\omega)=\beta_0+\beta_1 (\omega-\omega_0)+\frac{\beta_2}{2}(\omega-\omega_0)^2+ \ldots, 
 \label{eq:taylorbeta}
\end{align}
 where $\beta_n=\frac{\partial^n \beta}{\partial \omega^n}|_{\omega_0}$.

 The nonlinear response of the medium is governed by the susceptibility tensor $\chi^{(3)}$, which we assume to be real for a lossless medium. We assume an instantaneous nonlinear response, i.e. we neglect the time dependence of $\chi^{(3)}$. This amounts to neglecting contributions from (slow) molecular vibrations, also known as the Raman effect \cite{Agra-NL}, over the fast electronic response \cite{Boyd}. This approximation is reasonable for pulse lengths down to $\sim 100\,{\rm fs}$ or short propagation distances.  For the nonlinear polarization, these approximations implemented in (\ref{eq:polar}) then imply
 \begin{align}
 P_{\mathrm{NL}}(z,t)=\epsilon_0 \chi^{(3)}_{xxxx} E^3(z,t), 
 \label{eq:nondelayedchi3}
\end{align} 
where $\chi^{(3)}_{xxxx}$  is the relevant tensor element if all fields are directed in the $x$-direction\footnote{The instantaneous response implies that $\chi^{(n)}$ in (\ref{eq:polarizationextension}) is proportional to the $\delta$-function, i.e. has different units compared to $\chi^{(3)}_{xxxx}$.}. As for the linear refractive index, we define an effective nonlinear coefficient in the fiber, $\chi_{\mathrm{eff}}^{(3)}\approx\chi^{(3)}_{xxxx}$.
Note that the optical Kerr effect, which manifests in the intensity-dependent refractive index, is only part of the total nonlinear response described by $\chi_{\mathrm{eff}}^{(3)}$.

We now carry out the separation according to Eq. (\ref{eq:separation}). As mentioned above, the transverse  profile $F(x,y)$ is responsible for non-bulk, effective quantities, but is otherwise assumed constant along the fiber. We obtain a scalar equation similar to (\ref{wave3}), but without the transverse coordinates: 
\begin{align}
\partial^2_z E(z,t) + \hat{\beta}^2(i\partial_t)E(z,t) - \frac{1}{c^2}\partial^2_t \chi^{(3)}E^3(z,t)=0, 
 \label{eq:realpropagationequation}
\end{align}
where the linear response is described by (\ref{eq:susceptibilitytoindex}) and the expansion (\ref{eq:taylorbeta}) has been used.  Although this equation seems complex due to the explicit $i$, it is entirely real as $\hat{\beta}^2$ is a real even function and the electric field is the real electric field.

We proceed to complexify the electric field and transform the propagation equation into a frame moving with the light in the fiber. This modification, which does not involve further approximations,  will allow us to read off important physical effects related to analogue gravity.
We write the electric field as:
\begin{align}
E= \frac{1}{2 \sqrt{A_{\mathrm{eff}}\epsilon_0 c \hat{n}_{\mathrm{eff}} }}(E_c+c.c.),
 \label{eq:complexification}
\end{align}
where $E_c=E_c(z,t)$ is the complex electric field. The effective mode area $A_{\mathrm{eff}}$\footnote{The effective mode area is defined from the transverse mode profile $F(x,y)$. Its precise definition can be found in, e.g. \cite{Agra-NL}.}  and the other normalization terms allow us to express $E_c$ in units of $W^{1/2}$ rather than the $V/m$ of $E$, relating the field to the optical power inside the fiber.  The resulting wave equation for $E_c$ is
\begin{align}
\partial^2_z E_c(z,t) + \hat{\beta}^2(i\partial_t)E_c + \frac{2}{3c} i \partial_t \hat{n}_{\mathrm{eff}} \hat{\gamma}(E^3_c+3 E_c^2 E^*_c)=0, 
 \label{eq:complexpropagationequation}
\end{align}
with
\begin{align}
\hat{n}_{\mathrm{eff}}(i\p_t)\, \hat{\gamma}(i\p_t)=\frac{3\chi^{(3)} }{8\,A_{\mathrm{eff}}\epsilon_0 c^2 \hat{n}_{\mathrm{eff}}}i\p_t. 
 \label{eq:definitiongamma}
\end{align}
Note that the complexification introduces an arbitrary imaginary part in $E_c$ first, which we then can constrain such that there is no complex conjugate term in  (\ref{eq:complexpropagationequation}). We can see that the nonlinear term of the equation splits into a third power term describing third harmonic generation and a second term describing `self-phase modulation'. 
If we momentarily disregard the third harmonic term, we can compare equation (\ref{eq:complexpropagationequation}) to the Helmholtz equation (\ref{Helm.E}) and read off that
\begin{align}
\frac{\omega^2}{c^2} n^2(z, \omega)= {\beta}^2(\omega)+2 \frac{\omega\, n_{\mathrm{eff}}(\omega)  }{c }\gamma(\omega) \widetilde{|E_c|^2}(z,\omega), 
 \label{eq:nonlinearindex1}
\end{align}
where we Fourier transformed the right hand side and $\widetilde{|E_c|^2}(z,\omega)$ is the optical power density.
Thus, for small nonlinear fiber coefficients $\gamma(\omega)$:
\begin{align}
n(z, \omega)\approx n_{\mathrm{eff}}(\omega) +\frac{c \gamma(\omega)}{\omega} \widetilde{|E_c|^2}(z,\omega),
 \label{eq:nonlinearindex2}
\end{align}
with 
\begin{align}
\gamma(\omega)=\frac{3 \omega \chi^{(3)}}{8 c^2 A_{\mathrm{eff}}\epsilon_0 n^2_{\mathrm{eff}}(\omega)}
 \label{eq:definitiongamma2}
\end{align}
is obtained as an intensity dependent refractive index. This change of index under the influence of light is the optical Kerr effect, which means that intense light can bend light, the principle of optical analogues. 

We now separate off a fast oscillating phase, typically around an optical carrier frequency, by $E_c = E_0 e^{i(\beta_0 z-\omega_0 t)}$, which leads to the following substitution rules:
\begin{align}
\partial_z &\rightarrow (\partial_z+i\beta_0)\quad
\partial_t \rightarrow (\partial_t-i \omega_0).
 \label{eq:carriersubrules}
\end{align}
Next, we transform into coordinates that move with the light at speed $v$:
\begin{equation}
\begin{array}{ll}
\tau= t-z/{v} \equiv t-\beta_1 z\quad \quad
&\zeta = z\\
\partial_t = \partial_\tau\quad \quad
&\partial_z = \partial_\zeta-\beta_1 \partial_\tau.
\end{array}
 \label{eq:galileantransform}
\end{equation}

Note that any moving frame such as an (inertial) Lorentz frame is equally appropriate, as this is merely a coordinate transformation. The pulse is centered at the retarded time $\tau=0$. Formally, in the analogue gravity context in the following sections, we will regard $\tau$ as space and $\zeta$ as time, lead by the analogy to the relativistic wave equations.
Finally, we include 2nd and higher order dispersion in an expansion $D$:
\begin{align}
D &= \sum_{n=2}^{\infty} \frac{\beta_n}{n!} (\omega-\omega_0)^n \quad \leftrightarrow \quad\beta(\omega) = \beta_0+\beta_1 (\omega-\omega_0)) + D \label{eq:D}\\
\hat{D} &= \sum_{n=2}^{\infty} \frac{\beta_n}{n!} (i\partial_\tau)^n\quad \leftrightarrow \quad \hat{\beta}(i \partial_\tau +\omega_0) = \beta_0+\beta_1 i \partial_\tau + \hat{D}.
 \label{eq:Dhat}
\end{align}
With these modifications we write the propagation equation (\ref{eq:complexpropagationequation}) in terms of $E_0$ and obtain in the moving frame:
\begin{align}
\left[ \hat{D}-i \partial_\zeta + 2(\beta_0+\beta_1 i \partial_\tau)\right]\left[ \hat{D}+i \partial_\zeta )\right]E_0+\frac{2(i\partial_\tau+\omega_0)}{3c}\hat{n}_{\mathrm{eff}}\, \hat{\gamma}\left( E^3_0 e^{2i(\beta_0 z - \omega_0 t)}+3|E_0|^2 E_0 \right) &=0.
 \label{eq:propagationequationtime}
\end{align}
This is the propagation equation for intense light in the fiber. As the major approximation, we neglected the Raman effect in the nonlinear response in equation (\ref{eq:nondelayedchi3}). We retain the second spatial derivative as well as higher order dispersion and we have avoided the slowly varying envelope approximation. In the context of analogue gravity, where often we have to handle very wide bandwidth fields, this is very useful. These narrowband approximations are usually found in the literature \cite{Agra-NL}.  
The equation can be Fourier-transformed in time and space ($\partial_\zeta \rightarrow ik, \quad \partial_\tau \rightarrow -i(\omega-\omega_0)$) and becomes:
\begin{align}
\underbrace{\left[ {D}+k + 2(\beta_0+\beta_1 (\omega-\omega_0)\right]\left[ {D}-k )\right]\tilde{E}_0}_{\mbox{linear, homogeneous oscillator}}+\underbrace{\frac{2\omega}{3c}n_{\mathrm{eff}}\,\gamma\, \mathbfcal{FT}\left[ E^3_0 e^{2i(\beta_0 z - \omega_0 t)}+3|E_0|^2 E_0 \right]}_{\mbox{coupling}}&=0,
 \label{eq:propagationequationfrequency}
\end{align}
where $\mathbfcal{FT}$ denotes the Fourier transform in space and time. We can see that the left part of (\ref{eq:propagationequationfrequency}) describes the dispersive, linear field evolution with the square brackets giving the dispersion relation, whereas the right part describes the nonlinear interaction, which acts to modulate and couple the linear modes.

The third harmonic term is not generally phasematched, but might be responsible for a rather low photon noise background.

\subsection{The nonlinear Schr\"odinger equation}

In this section we demonstrate how to reduce Eq. (\ref{eq:propagationequationtime}) to the well known nonlinear Schr\"odinger equation using further approximations. We then briefly look at simple, but paradigmatic solutions to this equation. 

We start with neglecting the first of the nonlinear terms, the third harmonic term. This term generates a wave at three times the frequency compared to its generating field $E_0$, which propagates, in most fibers, with a slower phase velocity than $E_0$, i.e. the waves are not phasematched and no significant wave amplitude builds up. 
Using Eq. (\ref{eq:D}) in (\ref{eq:propagationequationfrequency}), we obtain 
\begin{align}
\left[ 2 {\beta}(\omega)-({D}-k)\right]\left[ {D}-k \right]\tilde{E}_0+2 {\beta}(\omega){\gamma}(\omega)\widetilde{|E_0|^2 E_0}  =0.
 \label{eq:FTpropagationequationgamma}
\end{align}
The effect of the Kerr nonlinearity in a fiber can be similar in size to that of the group velocity dispersion $D$, which is of 2nd order in $(\omega-\omega_0)$ in Eq. (\ref{eq:taylorbeta}), but is much smaller than the wavenumber $\beta(\omega)$ or $\beta_0$ itself, which is of zeroth order. Therefore, in Eq. (\ref{eq:FTpropagationequationgamma}) we can neglect $D$ and $k$ in the first bracket,
 if $(\omega-\omega_0)$ is of a limited size justified  by the expansion (\ref{eq:taylorbeta}). In fact, for a backpropagating wave ($\beta(\omega)<0$) this approximation is broken and thus we expect unidirectional propagation hereafter. After removing the common $\beta(\omega)$ factor and another space-time Fourier transform we arrive at:
\begin{align}
\left[ \hat{D}+i \partial_\zeta \right]E_0+\hat{\gamma}(i\partial_\tau+\omega_0)|E_0|^2 E_0  =0,
 \label{eq:GNLSE}
\end{align}
which is known as the generalized nonlinear Schr\"odinger equation (GNLSE) without the Raman delayed response \cite{wood1988}. Finally, neglecting all but the lowest orders in $\hat{D}$ and $\hat{\gamma}$, we arrive at the nonlinear Schr\"odinger equation (NLSE):
\begin{align}
\left[ -\frac{\beta_2}{2}\partial^2_\tau+i \partial_\zeta \right]E_0+{\gamma_0}|E_0|^2 E_0  =0,
 \label{eq:NLSE}
\end{align}
where $\gamma_0=\gamma(\omega)|_{\omega=\omega_0}$. The NLSE is the principal approximation for nonlinear wave evolution in optical fibers \cite{Agra-NL, Boyd}. It is an integrable equation with stationary pulse shapes known as solitons.

\subsection{Solutions to the nonlinear Schr\"odinger equation}

\begin{figure}[H]
\begin{center}
\includegraphics[scale=0.85]{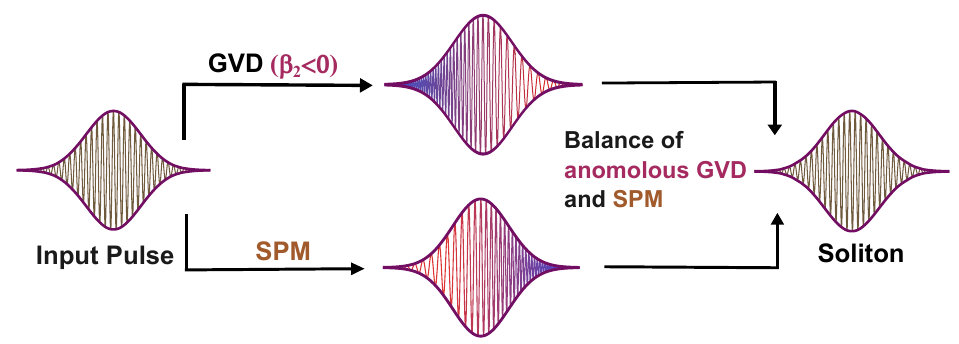} 
\caption{Soliton formation: The effects of group-velocity dispersion and nonlinearity conspire to generate opposite chirps that cancel to produce an unchirped stable pulse.}  
\label{fig:solitonformation}
\end{center}
\end{figure}

We will now look at some elementary analytical solutions of the NLSE (\ref{eq:NLSE}). The evolution of the field, $\partial_\tau E_0$, is described by two terms: the nonlinear term $\gamma_0|E_0|^2E_0$ and the dispersion term $-\frac{1}{2}\beta_2\p_T^2 E_0$. The former is responsible for the phenomenon of \textit{self-phase modulation (SPM)}---that is, the  field broadens spectrally by acquiring an intensity-dependent temporal phase shift given by $\Delta \phi=\frac{{|E_0|^2}}{A_{\mathrm{eff}}}\gamma_0 \zeta$, cf. Eq. (\ref{eq:nonlinearindex2}),---while the latter, which is referred to as \textit{group velocity dispersion (GVD)} causes the field to broaden temporally as it propagates along the fiber due to a spectral phase shift $\Delta \phi = \frac{\beta_2}{2} \Omega^2 \zeta$.  In other words, dispersion advances, e.g. blue ($\omega>\omega_0$) frequency components to propagate faster, i.e. they would show up in the front of a pulse and vice versa for red components. The nonlinearity, on the other hand, creates a decreasing phase change in the front of the pulse, lowering the frequencies of waves in the front  and vice versa in the back. This is illustrated in figure \ref{fig:solitonformation}.

The simplest solution to the NLSE is a field uniform in intensity:
\begin{align}
E_0 = \sqrt{P_0}\,\,e^{i\gamma_0 P_0 \zeta}.
 \label{eq:cwsolution}
\end{align}
This simple solution is interesting as it is an unstable solution in a fiber of negative -- anomalous -- group velocity dispersion ($\beta_2<0$). In this case, small fluctuations in the field are amplified and spontaneously pulses are formed, as the field `bunches up'. This amplification is a parametric process of four-wave mixing of the initial field and two equally spaced sidebands \cite{Agra-NL}. It is also known as modulation instability and demonstrates that the field experiences an attractive interaction in anomalous dispersion, something we will get back to in section \ref{sec:blackholeoscillations}.

In the just mentioned anomalous-dispersion regime, the fiber can also support wave packet (pulsed) solutions that can travel with a preserved envelope shape $E_0(\zeta,\tau)$. These solutions are called optical \textit{solitons} and find many applications, e.g. in lasers. In the context of analogue gravity, this is interesting as these `static' pulses modify the propagation speed of other light in the fiber, effectively creating a non-flat, but static  spacetime in 1+1 dimensions. We discuss this topic in the next section.

The preserved pulse profile is a hyperbolic secant, leading to a solution \cite{Agra-NL}:  
\begin{equation}
E_0(\tau,\zeta)=\sqrt{P_0}\,\text{sech}\left({\frac{\tau}{T_0}}\right)e^{i\gamma_0 P_0 \zeta/2}, \label{soliton}
\end{equation}
where $P_0$ is the power of the intense pulse at its peak and $T_0$ is a parameter that determines its width. This solution is called the `fundamental soliton'. For this to solve Eq. (\ref{eq:NLSE}), we need anomalous dispersion and a peak power $P_0$ that balances dispersive and nonlinear effects:
\begin{equation}
\frac{1}{\gamma_0 P_0}=\frac{T^2_0}{|\beta_2|}, \label{eq:solitoncondition}
\end{equation}
which is also known as the soliton condition.
More generally, to study the interplay between SPM and GVD, one can define the ratio parameter
\begin{equation}
N\equiv \frac{\gamma_0 P_0 T_0^2}{|\beta_2|},
\end{equation}
which is called the soliton order.
The nonlinear effects are  apparent  when $N>1/2$, while dispersion dominates when $N< 1/2$. For $N=1$ (cf. Eq. (\ref{eq:solitoncondition})) the two effects balance and the solution to the NLSE is the \textit{fundamental soliton}.

\section{Analogy of the event horizon}
\label{sec:fibereventhorizon}
In the previous section, we derived propagation equations for an intense optical field in a fiber. We discussed that the balance of nonlinear and dispersive effects can give rise to stable solutions: solitons. The field intensity locally and instantaneously increases the index of refraction via the Kerr effect, according to $\delta n(z,\omega)=\frac{c \gamma(\omega)}{\omega} \widetilde{|E_c|^2}(z,\omega)$ (cf. Eq. (\ref{eq:nonlinearindex2})). Thus, a small change $\delta n(z,t)$ of the refractive index moves along the fiber at the speed of the soliton $v_s=\beta^{-1}_1(\omega_s)$. This change, which is constant in the frame co-moving with the soliton ($v=v_s$), is what we regard as our background analogue spacetime. In this section, we are interested in exploring the motion of an optical probe wave on this background. This probe is weak, such that it interacts only linearly with the fiber and the background. In fact, if it is in the quantum vacuum state, it may be converted to photon pairs by the analogue Hawking effect.

\subsection{Propagation equation for probe waves}
\label{ssec:propagation for probe waves}
We now extend the field propagating in the fiber to be a soliton and a probe wave and we focus on the evolution of the probe. We start with the general wave equation (\ref{eq:propagationequationtime}) with the ansatz $E_0=E_s+E_p$, where $E_s$ and $E_p$ are the soliton and probe temporal envelopes. Note that (\ref{eq:propagationequationtime}) does not assume the field to be of narrow bandwidth, which means that soliton and probe can have very different carrier frequencies $\om_s$ and $\om_p$. If we neglect the third harmonic term ($E^3_0$-term) 
and if we set $\beta_0=\omega_0=0$, we obtain in linear order in $E_p$:
\begin{equation}
\left[ \hat{\beta}^2(i \p_\tau)+(\partial_\zeta-\beta_1 \partial_\tau)^2\right]E_p+4\, \hat{\beta}(i\p_\tau)\, \hat{\gamma}(i\p_\tau)\,|E_s|^2 E_p=0, \label{eq:horizonperturbation1}
\end{equation}
where we have neglected a fast oscillating term $\sim E^2_s E^*_p$, which is not usually phasematched.
In what follows, it is convenient to write the wave equation for the vector potential $A$. With  $E=-\p_\tau A$ the wave equation writes
\begin{equation}
c^2 (\partial_\zeta-\beta_1 \partial_\tau)^2A_p
-\p_\tau \left[ \left(  \hat{n}_{\mathrm{eff}}^2  +4 \, \hat{n}_\mathrm{eff}\, n_2\,|E_s|^2/A_{\mathrm{eff}} \right) \partial_\tau A_p \right]=0. \label{eq:horizonperturbation2}
\end{equation}

 We also have used $\hat{\gamma}=\frac{n_2 i\p_\tau}{c A_{\mathrm{eff} }}$, where $n_2$ is the slope of the linear dependence of the refractive index on optical intensity (cf. Eq.  (\ref{eq:nonlinearindex2})).
In (\ref{eq:horizonperturbation2}), the term in the large round  brackets plays the role of the square of the total -intensity dependent- refractive index, i.e. for small nonlinearities we obtain:
\begin{equation}
  n_{\mathrm{total}}\approx \hat{n}_{\mathrm{eff}}   +2  n_2 |E_s|^2/A_{\mathrm{eff}}. \label{eq:nonlinearcrossindex}
\end{equation}
This result is expected in comparison with Eq. (\ref{eq:nonlinearindex2}) with the important difference that the index that the probe experiences from the soliton is twice larger, i.e. cross-phase modulation is twice stronger than self-phase modulation.

\subsection{Ray tracing}
\label{ssec:raytracing}
\begin{figure}[H]
\begin{center}
\includegraphics[scale=0.65]{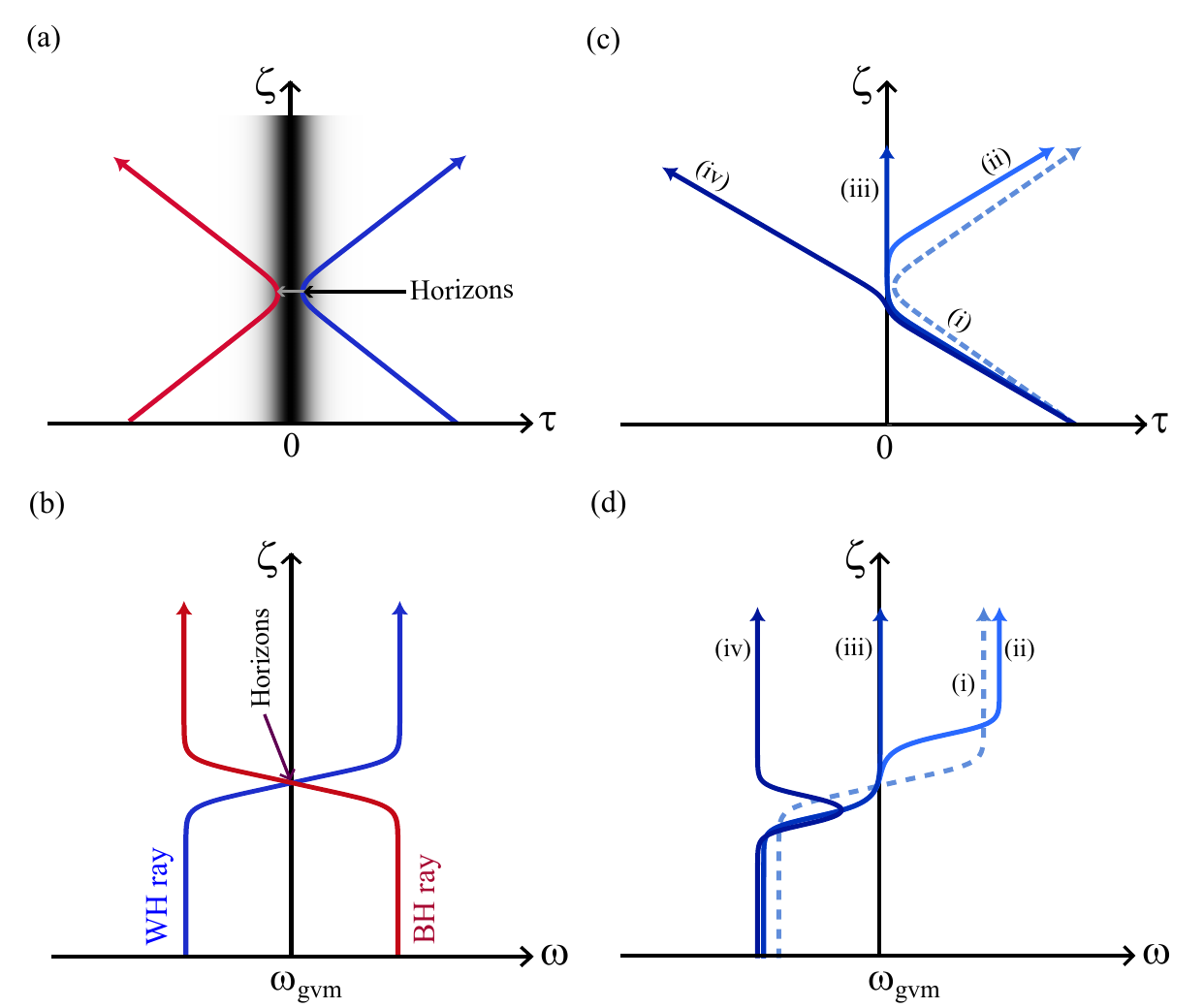} 
\caption{Rays interacting with a soliton: (a) Spatial domain: The grey shading represents soliton intensity for illustration. Rays that experience a black hole (red) and a white hole (blue) horizon at the front ($\tau<0$) and the back ($\tau>0$) of the soliton, respectively. (b) Frequency domain: Rays of (a) are blue shifting at the white hole (blue) and red shifting at the black hole (red).  (c-d) Spatial and frequency domain interactions for (i) the typical white hole interaction from (a,b), (ii) a faster ray entering further into the soliton, (iii) a ray riding on top of the soliton, (iv) a fast ray passing through the soliton. All rays propagate under normal dispersion. $\omega_{\text{gvm}}$: a frequency which is group velocity matched to the soliton.}  
\label{fig:raytracing}
\end{center}
\end{figure}

In order to understand how probe light governed by Eq. (\ref{eq:horizonperturbation1}) behaves around a soliton, we first consider rays in a geometric optics approximation. We assume a slowly $\tau$- varying nonlinear term, which is consistent with the JWKB approximation \cite{landau1987}. The ray trajectories will manifest the existence of event horizons. The ray tracing formalism can be found in, e.g. \cite{KipThorne}. We linearize $P_s(\tau)=|E_s|^2(\tau)$ and $\gamma(\omega)\approx\gamma_0+\gamma_1 (\omega-\omega_0)$ and neglect higher order derivatives in $\tau$ ($\p^2_\tau P_s, \gamma_2,\p_\tau P_s \gamma_1$,...).
We start by reverting Eq. (\ref{eq:horizonperturbation1}) to the laboratory frame using the substitution rules (\ref{eq:galileantransform}). We then insert a wave ansatz $E_p=e^{i(k z-\omega t)}$\footnote{$k$ is used independently here, not referring to $k$ from equation (\ref{eq:FTpropagationequationgamma}).} and approximate 
\begin{equation}
 \hat{\beta}(i \p_t) \hat{\gamma}(i \p_t) P_s(\tau)e^{i(kz-\omega t)} \approx \left[ \beta(\omega) \gamma(\omega) P_s(\tau)+i \p_\omega \left[ \beta(\omega) \gamma(\omega) \right] \p_t P_s(\tau)\right] e^{i(kz-\omega t)},   \label{eq:eikonalapprox}
\end{equation}
where $P_s(\tau)=|E_s(t-z/v, z)|^2$ is the soliton instantaneous power. This is neglecting higher derivatives of the slowly varying $P_s(\tau)$.  
The ensuing  expression for the  wave number $k$ is complex, but in the considered approximation, where $\p_\tau P_s$ is small, we arrive at the following real expression for the eikonal-approximated dispersion relation:    
\begin{equation}
-k^2+\beta^2(\omega)+4 \gamma(\omega) \beta(\omega) P_s(\tau) =0.
   \label{eq:eikonal}
\end{equation}
The ray tracing equations are extracted by applying the Hamilton equations with $\omega(k,z,t)$ as the Hamiltonian. We obtain the temporal changes in position $z$, frequency $\omega$ and wave number $k$ of the ray. Starting with deriving Eq.(\ref{eq:eikonal}) with respect to $k$ at constant $z$ and $t$, we obtain:
\begin{equation}
\frac{d}{dt}z=\frac{\p \omega}{\p k}=\frac{k}{\beta \beta_1 + 2 \gamma \beta_1 P_s+ 2 \gamma_1 \beta P_s}.
   \label{eq:raytracing1}
\end{equation}
Note that here we suppressed the dependence of $\beta, \beta_1$ and $\gamma$ on $\omega$ and of $P_s$ on $\tau$. The other two relations follow en suite:
\begin{equation}
\frac{d}{dt}vk =v\frac{d}{dt}k=\frac{-2\gamma \beta \, \p_\tau \!P_s}{\beta \beta_1 + 2 \gamma \beta_1 P_s+ 2 \gamma_1 \beta P_s}= \frac{d}{dt}{\omega}.
   \label{eq:raytracing23}
\end{equation}
It immediately follows from Eq. (\ref{eq:raytracing23}) that the frequency in the moving frame, $\omega'=\omega-vk$, is conserved, as expected for a stationary index profile. To simplify further, we can solve Eq. (\ref{eq:eikonal}) in our linear order and obtain:
\begin{equation}
k=\pm (\beta+2 \gamma  P_s),
   \label{eq:eikonallinear}
\end{equation}
which leads to the final form of the ray tracing equations:
\begin{equation}
\dot{z}=\pm \frac{1}{\beta_1+2 \gamma_1 P_s}\quad 
\dot{\omega}=v\dot{k}=\frac{-2\gamma_0\p_\tau \!P_s}{\beta_1+2 \gamma_1 P_s}.
   \label{eq:eikonalsolution}
\end{equation}
The sign choice indicates that rays can travel either direction in the fiber. A ray's velocity is determined  by the group velocity of the fiber $\beta^{-1}_1$, which depends on its frequency, and might be faster or slower than the soliton. A ray travels with a group velocity matched to the soliton at frequency $\omega_{\rm gvm}$. Ray trajectories are illustrated in Fig. \ref{fig:raytracing} (a-b). When approaching the soliton, the ray's frequency shifts and as a result its velocity changes. Then, for positive (negative) group velocity dispersion, the ray reflects off (accelerates through) the soliton, i.e. turning points on both sides of the soliton are generated in \textit{normal} dispersion (i.e. positive dispersion where the group velocity decreases with increasing frequency) as in Fig. \ref{fig:raytracing} (a-b). The light starts to reverse after the ray velocity reached the group velocity of the soliton. 
Figure \ref{fig:raytracing} (c-d) shows further ray trajectories for rays of various initial frequencies. Rays approaching the soliton ever so faster will penetrate further until they cross the soliton center, and the turning point disappears. 
\begin{figure}[H]
\begin{center}
\includegraphics[scale=0.75]{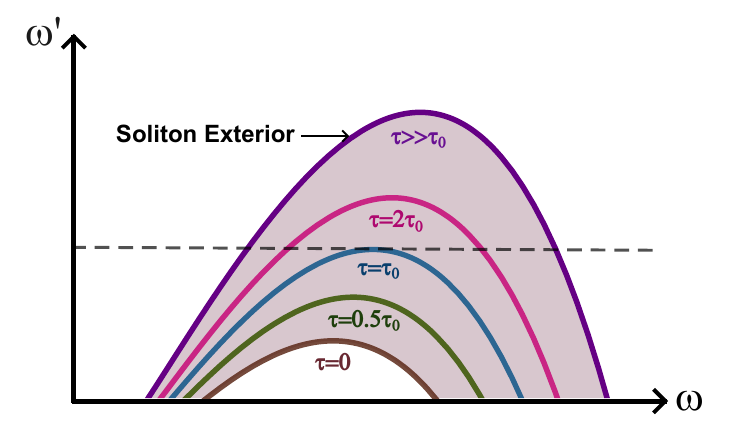} 
\caption{Dispersion curves outside and under the soliton in the co-moving frame: The co-moving frequency as function of $\omega$. $\tau$ is a position relative to the soliton, which is centered at $\tau=0$ and of width $\tau_0$.}  
\label{fig:dispersionmovingframe}
\end{center}
\end{figure}

This is further illustrated by considering the eikonal dispersion equation (\ref{eq:eikonal}) in the moving frame.  Fig. \ref{fig:dispersionmovingframe} shows the dispersion curves for different values of $\tau$, corresponding to different locations under the soliton. As the co-moving frequency is conserved, a valid ray with the example co-moving frequency indicated by the dashed line, lies on the intersection of this line with the dispersion curves, i.e. its frequency $\omega$ will change along this line. Because the slope of these dispersion curves indicates the co-moving velocity, the example ray can propagate with positive and negative velocity outside the soliton ($\tau \gg \tau_0$). Towards the soliton center ($\tau \rightarrow 0$), however, its velocity becomes zero (horizon, here at $\tau=\tau_0$) and no forward propagating rays exist beyond this point, leading to an effective `superluminal region' under the pulse. Thus, 
\begin{align}v\left[\beta_1(\omega)+2\gamma_1 P_s(\tau)\right]|_{\omega'}>1& \quad \mathrm{subluminal}\label{eq:groupvelocityhorizonsubluminal} \\
v\left[\beta_1(\omega)+2\gamma_1 P_s(\tau)\right]|_{\omega'}<1&\quad \mathrm{superluminal}
 \label{eq:groupvelocityhorizonsuperluminal}
\end{align}
are the conditions for the soliton to move sub- or superluminally\footnote{For superluminal regions the soliton is moving faster than any ray.}. 
Fig.\ref{fig:raytracing} illustrates these ray trajectories around the soliton. Note that the soliton may reflect or slow the ray similar to a repulsive potential, an aspect that will be important for the quasinormal mode description in section \ref{sec:blackholeoscillations}.

\subsection{Metric description of optical horizons}

An analogy between the propagation of light in fibers with curved spacetime emerges when we can find a suitable metric such that the wave equation of the optical probe field (\ref{eq:horizonperturbation2}) can be written with that metric. It would closely resemble the massless Klein-Gordon equation \cite{landau1987, Leonhardt2008}  
\begin{equation}
\p_\mu (\sqrt{-g}g^{\mu \nu}\,\p_\nu A_p)=0, \label{eq:KG.eqn}
\end{equation}
where $A_p$ is the vector potential of the probe field, $g^{\mu \nu}$ is the inverse metric tensor and $g<0$ is its determinant.  The indices $\mu$ and $\nu$ represent the coordinates $\zeta$ and $\tau$\footnote{In this notation $(\p_1, \p_2)=(\p_\zeta, \p_\tau)$.}, and we follow Einstein's index notation, according to which repeated indices indicate summation.

Our ansatz for a metric $g_{\mu\nu}=g_{\mu\nu}(\tau)$ is symmetric and depends on $\tau$ only, as the soliton background is stationary in $\zeta$.  We now use this metric in the Klein-Gordon equation (\ref{eq:KG.eqn}) and obtain
\begin{eqnarray}
g_{\tau\tau}\partial_{\zeta}^2 A_p
-2g_{\zeta\tau}\partial_\zeta\partial_\tau A_p
+(\partial_\tau g_{\zeta\zeta})\partial_\tau A_p
+ g_{\zeta\zeta}\partial^2_\tau A_p
-(\p_\tau g_{\zeta\tau})\partial_\zeta A_p \nonumber \\
-\frac{1}{2g}(\p_\tau g)\left[g_{\zeta\zeta}\partial_\tau A_p
-g_{\zeta\tau}\partial_\zeta A_p\right]=0. \label{eq:wave_equation_metric}
\end{eqnarray}
If we compare this equation with (\ref{eq:horizonperturbation2}), the coefficient of $\partial_\zeta^2 A_p$ implies $g_{\tau\tau}=c^2$, the coefficient of $\partial_\zeta\partial_\tau A_p$ implies $g_{\zeta\tau}=c^2\beta_1$, and the coefficient of $\partial_{\tau}^2A_p$ implies $g_{\zeta\zeta}=c^2\beta_1^2-n_{\text{total}}^2$. With these identifications, equation (\ref{eq:wave_equation_metric}) becomes:
\begin{equation}
c^2(\partial_\zeta-\beta_1 \partial_\tau)^2 A_p-\p_\tau \left( n^2_{\rm total} \p_\tau A_p \right) + \left( \p_\tau \ln n_{\rm total} \right) \left[ c^2 \beta_1 \p_z A_p + n^2_{\rm total} \p_\tau A_p \right]=0. \label{eq:wave_eqn_from_metric}
\end{equation}

Remarkably, the first two terms are identical to the wave equation (\ref{eq:horizonperturbation2}) of the probe field. The last term, however, introduces deviations, and its relevance depends on how strongly $n_{\text{total}}$ varies with $\tau$. In appendix A we show that under the condition
\begin{equation}
 \p_\tau \ln n_{\rm total}  \ll \omega, \label{eq:analogue_condition}
\end{equation}
the last term in equation (\ref{eq:wave_eqn_from_metric}) can be neglected as it is quadratic in $(\p_\tau \ln n_{\rm total})$. This approximation allows us to establish the analogy with a scalar field propagating in a curved spacetime. The associated metric reads:
\begin{equation}
g_{\mu\nu}=
\begin{pmatrix}
\frac{c^2}{v^2}-n_{\text{total}}^2(\tau) & \frac{c^2}{v}\\
\frac{c^2}{v} & c^2
\end{pmatrix}, \label{eq:metric}
\end{equation}
where we substituted $\beta_1=1/v$.

We can go one step further and ask when the probe wave equation (\ref{eq:horizonperturbation2}) is simulating the Klein-Gordon equation of a spin- and massless scalar field, i.e. the analogy is not only on the level of the modes, but includes the dynamics. Here we need, in addition, that (\ref{eq:horizonperturbation2}) is a second-order partial differential equation. This is the case in the absence of chromatic dispersion where $\hat{n}_{\text{eff}}(i\p_t) = \text{const.}$, i.e. $\hat{n}_{\text{eff}}$, and consequently $n_{\text{total}}$, does not contain derivatives.  The metric then describes a $1+1D$ Schwarzschild spacetime in the Painlevé–Gullstrand coordinate frame, with $\zeta$ being the proper time of free-falling observers and $\tau$ their radial coordinate. Such a geometry contains an event horizon at $\tau_{\text{h}}$, determined by $g_{\zeta\zeta}(\tau_\text{h})=0$.  Using the metric (\ref{eq:metric}), the location of a horizon is given by the solution to 
\begin{equation}
n_{\text{total}}( \tau_h)=\frac{c}{v}.  \label{eq:hor.cond}
\end{equation}
This condition states that the optical horizon in the nonlinear medium is generated at the location where the speed of the probe matches that of the soliton.  Having studied the generation of horizons in the nonlinear optical medium, we will now turn our attention to the Hawking effect.

\subsection{Hawking emission modes in a dispersive medium}
\label{ssec:hawkingmodes}

The optical horizon that reproduces the metric structure of a white- or black-hole horizon also leads to the spontaneous emission of photon pairs, even from the quantum vacuum state. This phenomenon indeed follows from the above analogy to white or black holes and the prediction of Steven Hawking \cite{Hawking74, Hawking75} for the radiation of black holes. We will now describe this process in the fiber in more detail. Compared to the astrophysical case examined by Hawking, optical media exhibit significant dispersion, calling for a re-examination of the permitted modes of propagation in the media. Therefore, we now present a medium dispersion model, an extension of the rather general microscopic `Hopfield model' \cite{hopfield1958}, as also explained in \cite{U46, jacquet2020PRA, trevisan2024}.

In this microscopic description, the dielectric medium is modeled as a set of harmonic oscillators representing the atomic dipoles of the medium. For probe fields with a wavelength much longer than the interatomic distance, the dipole-oscillators can be modeled in the continuous limit as fields.  Let these medium polarization fields be $P_1$, $P_2$, and $P_3$\footnote{The three fields describe a realistic medium sufficiently well in the transparency window, i.e. in the frequency range in which absorption is negligible.}. Let also $A$ be the scalar component of the light field vector potential $\bm{A}=\hat{x}A$, as introduced in section \ref{ssec:propagation for probe waves}. For simplicity, we drop the label ``$p$" from $A_p$. The Lagrangian density describing the model is \cite{finazzi2013,hopfield1958,fano1956}
\begin{equation}
\mathcal{L}=\frac{(\p_tA)^2}{8\pi c^2}-\frac{(\p_zA)^2}{8\pi}+\sum_{i=1}^3\left[\frac{(\p_tP_i)^2}{2\kappa_i\om_i^2}-\frac{P_i^2}{2\kappa_i}+\frac{1}{c}A\p_tP_i\right].\label{Hopfield.Lagrangian}
\end{equation}
There are three pairs $(\kappa_i, \om_i)$ characterizing three resonances in the medium. Parameter $\kappa_i$ is the elastic constant of oscillator $i$ while $\om_i$ is its eigenfrequency. We are concerned about frequencies of the electric field far away from these resonances and, thus, absorption losses are negligible. The first two terms of the Lagrangian describe the free electromagnetic field, the first two terms in the sum correspond to the dipole oscillators and the last term is their coupling. 
 
Deriving the equations of motion from the Lagrangian (\ref{Hopfield.Lagrangian}) and inserting a plane wave solution $e^{ikz-i\om t}$, we arrive at the dispersion relation \cite{finazzi2013}
\begin{equation}
 c^2k^2=\om^2\left(1+\sum_{i=1}^3\frac{4\pi \kappa_i}{1-\om^2/\om_i^2}\right). \label{disp.rel}   
\end{equation}
The dispersion is in the form of a Sellmeier relation. It accurately describes the off-resonant (linear) refractive index of the medium. 
\begin{figure}[H]
\begin{center}
\includegraphics[scale=0.65]{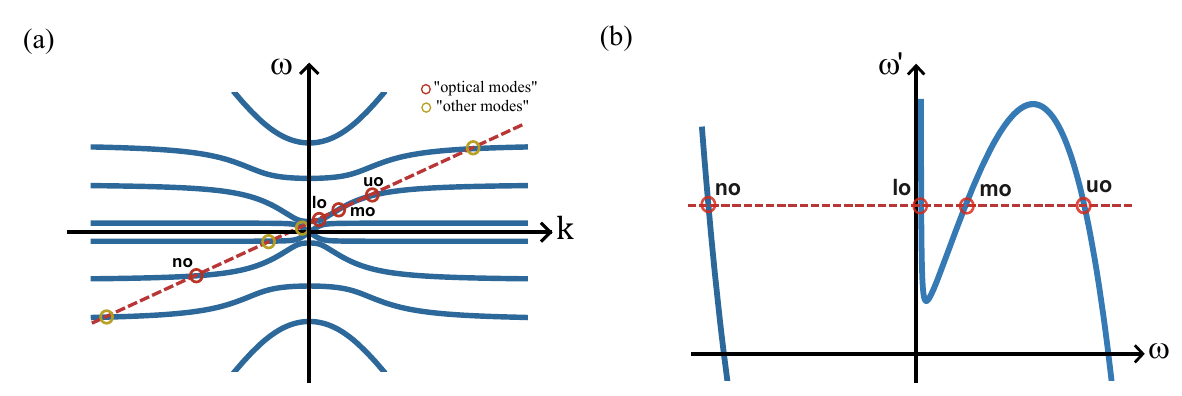} 
\caption{(a) Dispersion relation in the lab frame for a medium with three resonances according to the Hopfield model. The curves correspond to the square root of the RHS of (\ref{disp.rel}) while the straight line is a contour of a comoving frequency $\om^\pr=\om-vk$. Plane wave mode solutions of waves are indicated by red circles and we name the optical modes \textit{lower optical} ($lo$), \textit{medium optical} ($mo$), \textit{upper optical} ($uo$) and \textit{negative optical} ($no$). (b) Dispersion relation of the co-moving frequency.}  
\label{fig:dispersionrelation}
\end{center}
\end{figure}

Figure \ref{fig:dispersionrelation} shows the dispersion relation in lab (a) and co-moving (b) frames. The blue line indicates a contour of constant comoving frequency $ \om^\pr$. In section \ref{ssec:raytracing}, we had seen that waves interacting with the soliton strictly preserve this frequency. Therefore, a wave of $ \om^\pr$ is always composed of the plane waves defined by the intersection of the blue line with the dispersion curve. In view of the polynomial dispersion relation (\ref{disp.rel}), there are 8 such intersections, i.e. 8 possible `modes'. We are restricting our analysis to the optical modes, ``$no$", ``$Io$", ``$mo$" and ``$uo$", which are the only modes guided in the fiber. 

\begin{figure}[H]
\begin{center}
\includegraphics[scale=0.8]{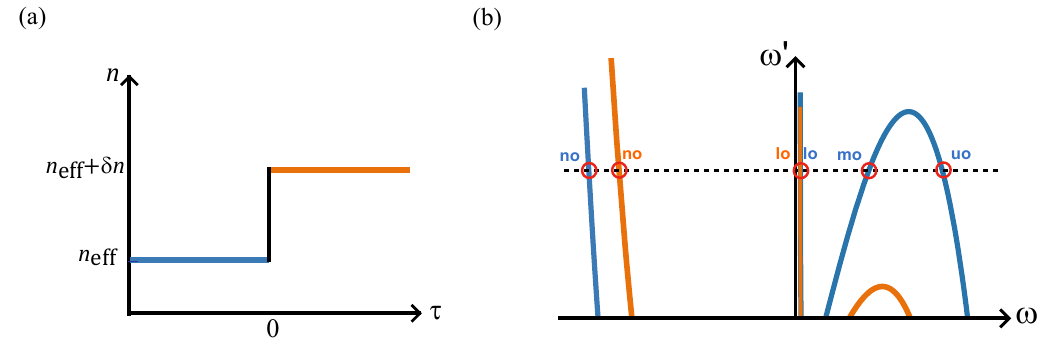} 
\caption{(a) The moving refractive index step as a simple model for an optical horizon. Note that $\tau >0$ is a region traveling behind $\tau<0$ in the fiber as $\tau = t-z/v$. (b) For a suitable $\om^\pr$ (dashed horizontal line), modes with positive and negative co-moving velocity do exist in front of the step  ($\tau<0$), but only left-moving modes for $\tau>0$. The optical horizon is located at co-moving frame position $\tau=0$.}
\label{fig:stepindex}
\end{center}
\end{figure}

To explicate particle creation by an optical horizon, let us adopt the simplest model allowing for a horizon, the moving refractive index step depicted in figure \ref{fig:stepindex}. An index step like this would not propagate unchanged through the fiber, like a soliton, but it allows for a simple, exemplifying quantum optical calculation of the mode conversion process. This configuration exhibits a single black-hole horizon.   
The step refractive index marks two regions: 1) $\tau>0$ where the refractive index $n_{\rm eff}+\delta n$ is raised (pulse interior) and 2) $\tau<0$ where the refractive index $n$ is normal (pulse exterior). The boundary $\tau=0$ is the black hole horizon as we explain in the following. In figure \ref{fig:stepindex} (b), we show the co-moving-frame dispersion relation in the two regions. 
The orange line is the dispersion curve in the interior region ($\tau>0$), while the blue line is the dispersion curve in the exterior. 
 Waves of a co-moving frequency $\omega'$ (below a cutoff) in the exterior consist of four solutions indicated in figure \ref{fig:stepindex} (b).
The slopes in these points differ in sign, meaning that some wave components may move towards $\tau=0$ while others may move away from it. So, while motion in both directions is possible outside the index front (the pulse), modes with left- or forward motion  cease to exist in the interior (orange line) \footnote{In this case the modes cease, because they become complex (evanescent).}.   
More concretely, in the exterior  region, modes $no$, $lo$ and $uo$ are right movers, i.e. propagation towards the horizon, while the mode $mo$ is a left mover, i.e. propagation away from the horizon. In the interior, modes $mo$ and $uo$ cease. Thus, $mo$ is a unique escaping mode, the \textit{Hawking mode}. In the exterior region waves may propagate in either direction, while in the interior, all waves propagate away from the horizon. It is precisely this structure that is analogous to black hole spacetimes. 

The particle pair generation is governed by the conversion of input modes into output modes. Incoming light is scattered into outgoing light, described in this stationary situation by the scattering matrix, which we will hence derive. Because the frequency in the moving frame, $\omega'$, is a conserved quantity, we can restrict our discussion to one arbitrary $\omega'$. 

The `scattering modes' are modes that are defined everywhere in space $\tau$. Asymptotically, for $\tau \rightarrow \pm \infty$ they may be identical to the modes indicated in Fig. (\ref{fig:dispersionrelation}) (a), but generally they are a $\tau$-dependent superposition of these. In order to find a mode basis for the scattering, we define modes that asymptotically correspond to only a single mode $\alpha \in \{no, lo, mo, uo, ...\}$. This gives us a set of eight modes on either side of the pulse, half thereof moving towards the pulse ("in"-modes) and half thereof moving away ("out"-modes). The "in" and the "out" modes form two bases for scattering modes. As we explain below, it is the change of basis that leads to a change of the vacuum state, i.e. the output vacuum is not the input vacuum and therefore contains particles.

Formally, we use the Lagrangian density (\ref{Hopfield.Lagrangian}) (in the moving frame\footnote{For simplicity, we keep denoting our coordinates $(\tau, \zeta)$. To be consistent with the literature in this section, these are however Lorentz transformed from the lab frame \cite{U46, jacquet2020PRA}.}) and define the canonical conjugate momenta of the fields $(A, P_1, P_2, P_3)$ by
\begin{align}
\Pi_A(T,\zeta)&=\frac{\delta \mathcal{L}}{\delta (\p_\zeta A)}, \label{conj.mom} \quad \quad
\Pi_{P_i}(T,\zeta)=\frac{\delta \mathcal{L}}{\delta (\p_\zeta P_i)}, \quad i=1,2,3.
\end{align}
The explicit expressions of the momenta are not relevant for the purposes of this section and can be found in \cite{U46, jacquet2020PRA}. 
Hence we combine the scattering modes in a vector
\begin{equation}
{V=\left(A, P_1, P_2, P_3, \Pi_A, \Pi_{P_1}, \Pi_{P_2}, \Pi_{P_3} \right)^\text{T}.} \label{V.vector}
\end{equation}

We complexify the real field $V$\footnote{For simplicity, we denote the real fields introduced and complexified from here on by the same symbols.}, which then exhibits a U(1) symmetry, which implies a conserved Noether current \cite{jacquet.thesis, cohen1977}.  We can use this to define a conserved scalar product on the space, the Klein-Gordon product
\begin{equation}
\left< V_{1}, V_{2}\right>=\frac{i}{\hbar}\int \!\!\! d\tau \, V^\dagger_{1}(\tau,\zeta) \,\left(\begin{array}{cc}0&\bm{I}_4\\-\bm{I}_4&0\end{array}\right)\, V_{2}(\tau,\zeta), \label{KG.product}
\end{equation}
 with $\bm{I}_4$ being the $4\times 4$ identity matrix. With the Klein-Gordon product we orthonormalize the scattering modes to \cite{finazzi2013}:
\begin{equation}
\left< V_{\alpha_1}(\om^\pr_1), V_{\alpha_2}(\om^\pr_2)\right>=\sgn(\alpha1)\, \delta_{\alpha_1 \alpha_2}\,\delta(\om^\pr_1-\om^\pr_2),
\end{equation}
where $\sgn$ stands for the sign of the norm of a mode. 

Now, the evolution in the optical modes of the refractive index front of Fig. \ref{fig:stepindex} is the following scattering process. There are three ingoing modes: $no$, $lo$, and $uo$ with group velocity towards the horizon from the left. After the interaction, the three outgoing modes are $no$, $lo$, and $mo$. While $no$ and $lo$ are outgoing to the right under the index front, $mo$ has a positive group velocity and is the `unique escaping mode'. Thus, the light moving outwards in the exterior of the black hole horizon is mode $mo$, which carries the Hawking emission and which we also call the \textit{Hawking mode}. Each Hawking photon is paired with a `partner' photon falling into the black hole. This mode has negative norm, to conserve the norm total, and therefore can be attributed to mode $no$.

 Formally, the scattering is a basis change from the in-basis to the out-basis\footnote{$S^T$ can be used instead of $S$, in which case the mode operators are not transformed by $S$, but by $S^T$. }:
 \begin{equation}
V^{\rm in}_{\alpha_1}=\sum_{\alpha_2} S_{\alpha_2\,\alpha_1}\, V^{\rm out}_{\alpha_2}. \label{eq:scatteringmatrix}
\end{equation}
In most cases the scattering matrix $S$ has to be obtained by a numerical solution of the equations of motion. For the step index, an analytical solution is possible \cite{jacquet2020PRA, jacquet2020SciPost}. As the main interaction happens among the optical modes $no$, $uo$, and $mo$ \cite{jacquet2020PRA}, we can simplify: 
\begin{equation}
\begin{array}{c@{\:\approx\:}c@{\:+\:}c}
    V^{\rm in}_{uo}& S^{}_{mo\,uo}\, V^{\rm out}_{mo}&S^{}_{no\,uo}\, V^{\rm out}_{no}  \\*[0.2cm]
    V^{\rm in}_{no}& S^{}_{mo\,no}\, V^{\rm out}_{mo}&S^{}_{no\,no}\, V^{\rm out}_{no} 
\end{array}
\qquad\mbox{or}\qquad
\begin{array}{c@{\:\approx\:}c@{\:+\:}c}
    V^{\rm out}_{mo}& S^{-1}_{no\,mo}\, V^{\rm in}_{no}&S^{-1}_{uo\,mo}\, V^{\rm in}_{uo} \\*[0.2cm]
    V^{\rm out}_{no}& S^{-1}_{no\,no}\, V^{\rm in}_{no}&S^{-1}_{uo\,no}\, V^{\rm in}_{uo}.
\end{array}
\label{eq:modetransformation}
\end{equation}

The mode mixing of two ingoing modes of opposite norm leads to an occupation of the Hawking mode and of the partner mode.

\subsection{Quantization and particle creation}
\label{ssec:paircreation}
Having derived the classical scattering formalism, we now can apply the canonical quantization procedure to calculate quantum expectation values. We promote the fields $A$, $P_i$, and their conjugate momenta to operators, satisfying the equal $\zeta$-time commutation relations 
\begin{align}
[\hat{A}(\tau_1,\zeta),\hat{\Pi}_{A}(\tau_2,\zeta)]&=i\hbar\, \delta(\tau_2-\tau_1), \\
[\hat{P}_{i}(\tau_1,\zeta),\hat{\Pi}_{P_i}(\tau_2,\zeta)]&=i\hbar \,\delta(\tau_2-\tau_1),
\end{align}
with all other commutators being zero. Collectively, we combine all these fields and momenta to the operator vector $\hat{V}$, defined analogously as in (\ref{V.vector}). We expand the field vector in creation and annihilation operators using the ingoing mode basis. As in the previous section, we only include the modes of interest, here $uo$ and $ no$ :
\begin{equation}
\hat{V}=\int^{\infty}_0 \!\!\! d\om^\pr \left[ V^{\rm in}_{uo}(\om^\pr)\hat{a}_{uo}(\om^\pr)+V^{\rm in}_{no}(\om^\pr)\hat{a}^\dagger_{no}(\om^\pr)\right]+\text{H.c.}, \label{V.Win.expansion}
\end{equation}
where ``H.c." stands for Hermitian conjugate. Whereas $\hat{a}$ is the amplitude of positive norm modes, the amplitude of negative norm modes is $\hat{a}^\dagger$, because the scalar product is an anti-hermitian sesquilinear form \cite{Leonhardt2008}.

We now insert the mode expansion (\ref{eq:modetransformation}) into the field expansion (\ref{V.Win.expansion}) and sort the result with respect to the out modes $V^{\rm out}$ to obtain the expansion of $\hat{V}$ in terms of out modes. This yields the \emph{out} annihilation and creation operators in terms of the \textit{in} operators:
\begin{equation}
    \left( \begin{array}{c} \hat{a}^{\rm out}_{\rm mo}\\
    \hat{a}^{\rm out \dagger}_{\rm no} \end{array}\right)= S \left( \begin{array}{c} \hat{a}^{\rm in}_{\rm uo}\\
    \hat{a}^{\rm in\dagger}_{\rm no} \end{array}\right).
    \label{eq:operatortransformation}
\end{equation}
The scattering matrix is transforming the mode operators from the input to the output. 
If the input quantum state of the optical field is the vacuum state $\ket{0}$,  defined as $ \hat{a}^{\rm in}_{\rm uo}\ket{0}$$=$$\hat{a}^{\rm in}_{\rm no}\ket{0}$$=$$0$, the number of quanta leaving the horizon in the Hawking mode is
\begin{equation}
N^{\rm out}_{\rm mo}=\bra{0} \hat{a}^{\rm out \dagger}_{\rm mo}\hat{a}^{\rm out}_{\rm mo}\ket{0}=|S_{mo\, no}|^2,\label{eq:particle.creation}
\end{equation}
which can be verified by direct insertion of (\ref{eq:operatortransformation}).
This is the Hawking effect in the optical analogue. Output mode $no$ is equally populated and the combined state, for each $\omega'$, is a `two mode squeezed state', or Einstein-Podolski-Rosen state, the photon statistic of which is thermal \cite{leonhardt.quantumoptics}. If we now consider the output state for \textit{any} comoving frequency, it is useful to define a temperature parameter as in \cite{Finazzi2012} in terms of the number of output photons
\begin{equation}
k_B T_{\text{}}=\frac{\hbar \omega'}{\ln{(1+|S_{mo\, no}|^{-2})}}.\label{eq:temperature}
\end{equation}
If this temperature parameter is frequency-independent\footnote{In a dispersive system, this is never the case. However, in some systems, a low-frequency regime can be found in which approximate thermality exists  \cite{Unruh,Kranas_thesis, ABK, Berti:2024cut}. This could only be in the low-frequency limit of the spectrum and considering a medium with sufficiently strong nonlinearity to block modes of arbitrarily low frequencies.},  the entire spectrum is thermal, i.e. the photon number follows a Bose-Einstein distribution, which is easily seen by making relation (\ref{eq:temperature}) explicit in $|S_{mo\,no}|^2$.

Finally, recall that this simple step model described here manifests a discrete spacetime with a sudden change at the interface at $\tau=0$. This is a crude approximation to the event horizon of a black hole, and thus we are not able to extract a surface gravity to predict a Hawking temperature; nonetheless, it is integrable, allowing one to obtain analytical insights \cite{jacquet2020PRA}. In a more realistic setup, with a continuous refractive index change, a surface gravity can be extracted, which is proportional to $\partial_\tau n_{\text{total}}|_{\tau_{h}}$, out of which follows a Hawking temperature $T$, as for example in \cite{Leonhardt2008, Belgiorno2011, Unruh, Kranas_thesis}. 
Up to this point, this section has established the connection between causal barriers and spontaneous particle creation. This connection is the defining feature of the Hawking effect.

\subsection{Mode conversion at the soliton: resonant radiation and negative frequencies}
In this section we will take a step back from analogue Hawking radiation and will take a brief look at other mode conversions present around the pulse traveling along the fiber. 
\begin{figure}[H]
\begin{center}
\includegraphics[scale=0.5]{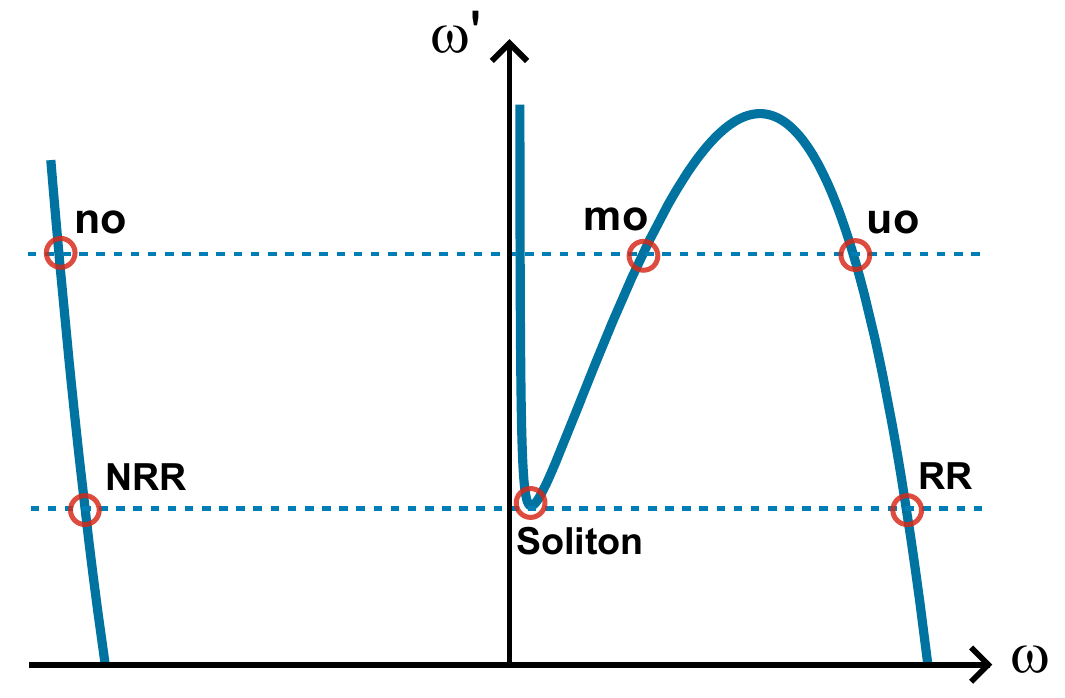} 
\caption{Two scenarios for the conversion of modes: Hawking effect (upper contour) and resonant radiation (lower contour).}  
\label{fig:conversions}
\end{center}
\end{figure}
In Fig. \ref{fig:conversions} we show the dispersion curve again in the moving frame of the soliton pulse at a position $\tau$ far from the pulse where the dispersion is independent of the pulse intensity. Any scattering of waves at the soliton conserves the comoving frequency $\om'$, i.e. incoming waves can only scatter into a discrete set of modes.  There are two contours of the comoving frequency shown in Fig. \ref{fig:conversions}:
The upper contour describes horizon physics studied before: an ingoing mode ($mo$) impinges on the back slope of the soliton, the white hole horizon, scatters to two modes ($uo, no$) of opposite norm, both falling behind the soliton\footnote{The fourth mode  involved is the negative norm input mode $no$ on the other side of the horizon, not shown in Fig. \ref{fig:conversions}. }. It can also describe scattering of mode $uo$, which is incoming at the front of the soliton, the black hole horizon, and scatters to two modes ($mo, no$) of opposite norm, one falling into the soliton and one escaping towards the front. These mode couplings are the same as in the Hawking effect as described in section \ref{ssec:hawkingmodes}; they lead to the spontaneous creation of photon pairs as in section \ref{ssec:paircreation}. However, for `bright' input modes that are not in the vacuum state, this is a classical mode conversion. This frequency conversion has been observed repeatedly, e.g. \cite{efimov2005,Leonhardt2008,choudhary2012,Tartara2012, drori2019}. For the black hole, also the negative norm output mode has been identified \cite{drori2019}. 
\begin{figure}[H]
\begin{center}
\includegraphics[scale=0.40]{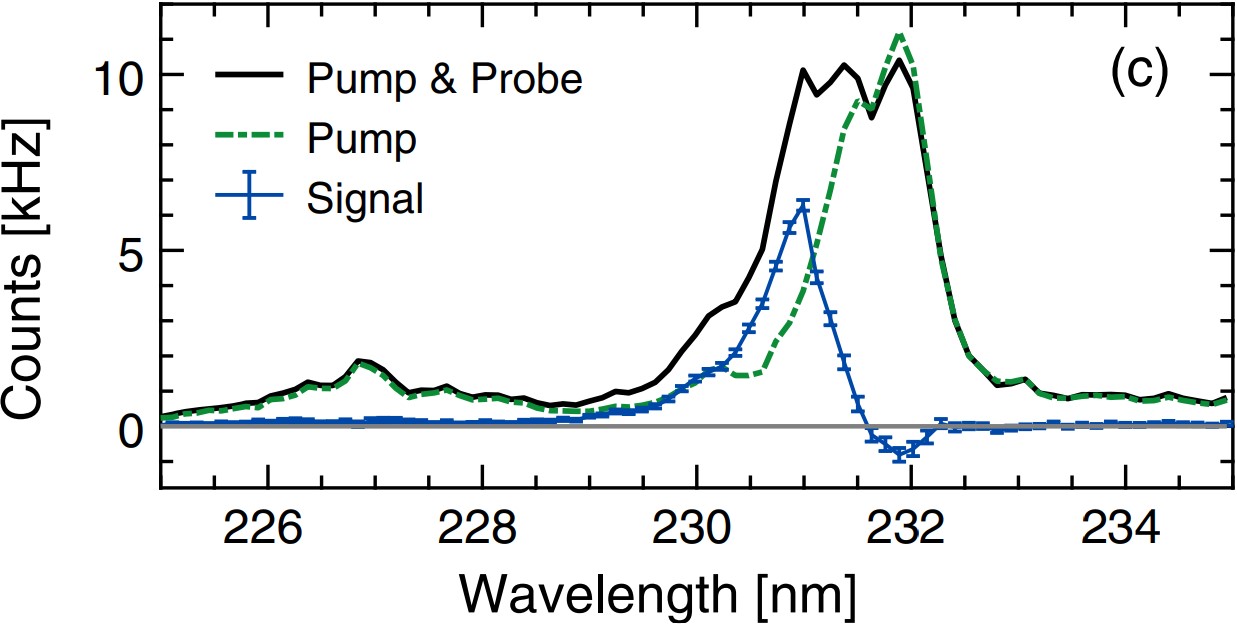} 
\caption{Spectrum  of the negative norm mode generated in the fiber-optical analogue Hawking effect (blue trace, which is the difference between the infrared probe in a bright state (black) and in the vacuum state (green)). \cite{Leonhardt2019}.}  
\label{fig:drori}
\end{center}
\end{figure}
Figure \ref{fig:drori} shows a spectrum of the mode $no$, which lies at high frequencies and short wavelengths in the ultraviolet part of the lab frame spectrum. The coupling is generally weak and the signals are feeble and hard to observe. An interesting aspect is the fact that the mode coupling is linear, i.e. it is independent of the intensity of light in these modes. The cross-phase modulation of these waves, which is necessary to create the lab frequency shift, is mediated by the soliton, which acts like a moving mirror.

The lower contour in Fig. \ref{fig:conversions} describes again scattering between the same modes. It is different though because the modes $lo$ and $mo$ coincide with each other and the soliton at the minimum of the dispersion curve, indicating that the soliton is standing still in the co-moving frame. Therefore, this mode is always highly excited and its photons can scatter to modes $uo$ and $no$. This is referred to as dispersive wave generation, resonant radiation, or fiber optic Cherenkov radiation in the literature \cite{akhmediev1995, skryabin2005, tartara2003, cristiani2004, herman2009, rubino2012, mclenaghan2014, menyuk1986, dudley2006, liu2015}. The process is readily understood by looking at the nonlinear wave equation, the Fourier transform of (\ref{eq:FTpropagationequationgamma}). Similar to section \ref{ssec:propagation for probe waves} we perturb with the ansatz $E_0=E_s+E_p$ and obtain in linear order:
\begin{align}
\left[ 2 {\beta}(i \p_\tau+\omega_0)-(\hat{D}+i\p_\zeta)\right]\left[ \hat{D}+i\p_\zeta \right]{E_p}+2 {\beta}(i \p_\tau+\omega_0){\gamma}(i \p_\tau+\omega_0)(2|E_s|^2 E_p+E_s^2 E^*_p)  =0.
 \label{eq:propagationequationRR}
\end{align}
If initially $E_p$ is a single frequency mode, the first term can oscillate at a single frequency. The second term, however, is nonlinear and therefore oscillates with an infinitely wide spectrum, which leads to the excitation of an ever so wider frequency spectrum in the first term as well, until the terms balance. In the frequency domain this is expressed in terms of the perturbation wave number $k_p$: 
\begin{align}
\left[ 2 {\beta}(\omega)-({D}-k_p)\right]\left[ {D}-k_p \right]\tilde{E}_p+2 {\beta}(\omega){\gamma}(\omega)\mathbfcal{FT}\left(2|E_s|^2 E_p+E^2_s E^*_p\right)  =0.
 \label{eq:FTpropagationequationRR}
\end{align}
Here we see that there are two resonance conditions for the oscillation of this field, namely if the first or second factor vanishes.  The perturbation field will have to increase to keep the equation balanced at that frequency, as the driving field is broadband. The conditions are:
\begin{equation}
\begin{array} {c@{\quad\quad}c}
D(\omega)=k_p & D(\omega)=k_p+2{\beta}(\omega),
 \label{eq:resonancecondition}
\end{array}
\end{equation}
where $D$ is the higher order dispersion defined in (\ref{eq:D}). With this definition it is easy to show that the resonances occur at:
\begin{equation}
\begin{array} {rl@{\quad\quad}rl}
 \omega-v \beta(\omega)&=\omega_0-v (\beta_0+k_p) &\omega+v \beta(\omega)&=\omega_0-v (\beta_0+k_p),\\
\omega_p'&=\omega_{0 p}'&\widebar{\omega_p'}&=\omega_{0 p}',
 \label{eq:RRconditions}
\end{array}
\end{equation}
which can be interpreted in the following way:
 $\omega_{0 p}'$ is the comoving frequency of a perturbation riding on the soliton, i.e. $k_p$ is the wavenumber shift due to the soliton cross-phase modulation. For a fundamental soliton, its nonlinear phase shift is $k_s = \gamma P_0/2$ and thus the perturbation experiences a nonlinear phase shift of $k_p=2 k_s=\gamma P_0$. The resonant wave perturbation, however, will leave the soliton into the medium without nonlinear phase contribution and its comoving frequency is $\omega_p'$. $\widebar{\omega_p'}$ on the other hand is a wave that propagates in the opposite direction and is therefore not phasematched. Equation (\ref{eq:RRconditions}) indicates that the process conserves momentum in the lab frame (phasematching), but energy (frequency) is conserved in the moving frame. It explains why the process can happen, although is seems energetically unfavorable as the photon energy in the lab frame increases: the moving frame energy is conserved and thus no energy is required for the conversion. 
 \begin{figure}[H]
\begin{center}
\includegraphics[scale=0.75]{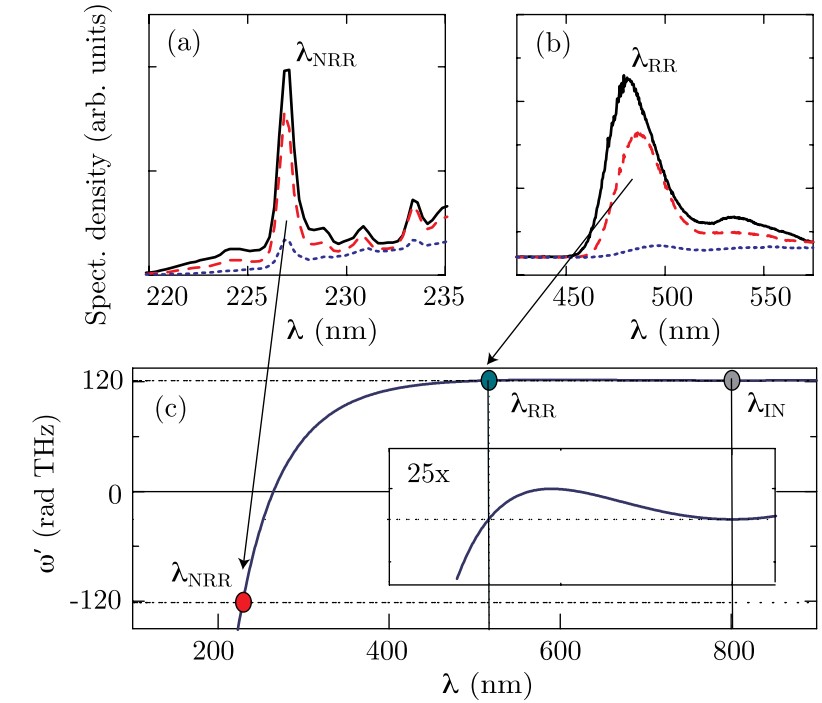} 
\caption{NRR experiment: (a-b) Measured spectra in the UV and visible region for three different input energies (solid, dashed, dotted lines). (c) Fiber dispersion relation with the predicted positions of the RR and NRR spectral peaks\cite{rubino2012} (Note the alternative sign convention: $\omega$ and $\lambda$ strictly positive and $\omega' \gtrless 0$.). }  
\label{fig:RR+NRR}
\end{center}
\end{figure}
 Figure \ref{fig:RR+NRR} shows a dispersion diagram identifying the resonant radiation (RR) and its negative frequency partner wave (NRR), together with the experimentally observed spectra. 
The resonant radiation process exemplifies how the physics of analogue gravity and the feeble Hawking effect can lead to processes of significant strength. It can be used to transform ultrashort pulses from e.g. the near infrared into the visible part of the spectrum with very high efficiency \cite{petty2020}.

\section{Oscillations of black holes}
\label{sec:blackholeoscillations}
We will now momentarily turn our attention back to relativistic black holes. We will explore oscillations of black holes and we will show how to create `analogue black hole oscillations' using solitons in optical fibers in what will be an entirely different black hole analogy compared to section \ref{sec:fibereventhorizon}. The recent interest in black hole oscillations \cite{cardoso2009, dreyer2003, gossan2012}, among others, stems from the discovery of gravitational waves in the LIGO collaboration \cite{abbott2016}. A black hole oscillation analogy is desirable to investigate black hole phenomena further that are hard to observe in space, such as the black hole spectral instability \cite{jaramillo2021, al2022scattering, cheung2022,
torres2023} or nonlinear effects \cite{mitman2023, cheung2023}.   

\subsection{Black hole quasinormal mode spectrum}
\label{sec:blackholespectrum}
\begin{figure}[H]
\begin{center}
\includegraphics[scale=0.65]{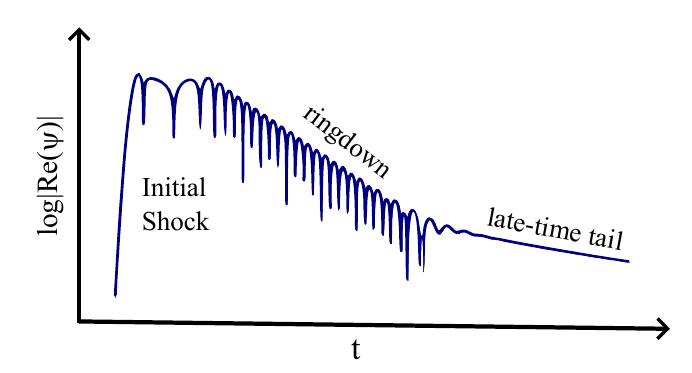}
\caption{Numerical ring down simulation of a Schwarzschild black hole \cite{ Kokkotas1999}.}
\label{fig:bhringdown}
\end{center}
\end{figure}

Black holes are stable solutions of spacetimes described by the Einstein equations \cite{MTWGravitation}. For example, the first black hole solution discovered is the Schwarzschild black hole with the metric
\begin{equation}
ds^2=-\left(1-\frac{r_s}{r}\right) dt^2+\left(1-\frac{r_s}{r}\right)^{-1} dr^2 + r^2 d\Omega,
\label{eq:schwarzschildlineelement}
\end{equation}
where $ds$ is the line element and $r_s$ is the Schwarzschild radius of the event horizon and spherical coordinates are used \cite{schwarzschild1999}. The perturbation of a black hole thus is expected to perform oscillations around the stable configuration and we can consider it as a field. Figure \ref{fig:light ring} illustrates this situation for light around the Schwarzschild black hole. A photon (massless field) can orbit the black hole, keeping the distance to the black hole center and the horizon constant. The gravitational attraction of the black hole is in balance with the centrifugal force. Hence, this orbit is unstable, as perturbations of the trajectory will lead to either the field falling into the black hole or spiraling off to spatial infinity. There are no stable light rings or photon spheres in Einsteinian gravity, unlike e.g. in conformal gravity \cite{Reinosuke2025}.  Formally, the massless scalar field ($\Psi$) is a relativistic quantum field and thus obeys the Klein-Gordon equation (\ref{eq:KG.eqn}) with the metric described by equation (\ref{eq:schwarzschildlineelement}).  The spherical symmetry lends itself to the following separation ansatz \cite{konoplya2011}: 
\begin{figure}[H]
\begin{center}
\includegraphics[scale=0.5]{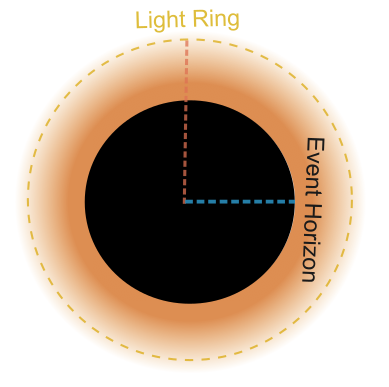}
\caption{Black hole with a light ring, a closed null-geodesic.}
\label{fig:light ring}
\end{center}
\end{figure} 
\begin{equation}
\Psi(t,r,\theta,\phi)=e^{-i \omega t}\,\,\frac{\psi(r)}{r}\,\, Y^m_l(\theta, \phi),
\label{eq:separationansatz}
\end{equation}
where $\omega$ is the frequency of oscillation and $Y^m_l$ are spherical harmonics. The resulting radial wave equation is of the Schr\"odinger type:   
\begin{equation}
\p^2_{r_*}\psi+\left(\omega^2-V(r)\right)\,\psi=0,
\label{eq:schrodingerequation}
\end{equation}
where $r_*$ is the `tortoise' coordinate:
\begin{equation}
\frac{dr}{dr_*}=\left( 1-r_s/r  \right).
\label{eq:tortoisecoordinate}
\end{equation}
The tortoise coordinate maps the radial coordinate from $(r_s,\infty)$ to $(-\infty, \infty)$, such that the event horizon lies at $-\infty$. The potential, generalized to fields of spin $s$, is: 
\begin{equation}
V(r)=\left( 1-r_s/r  \right)\left[ \frac{l (l+1)}{r^2}+\frac{(1-s^2) r_s}{r^3}  \right],
\label{eq:blackholepotential}
\end{equation}
where $l$ is the angular momentum and $s$ is the spin of the field. The light ring corresponds to an orbit on top of the potential barrier, which explains that it is unstable. 
\begin{figure}[H]
\begin{center}
\includegraphics[scale=1.0]{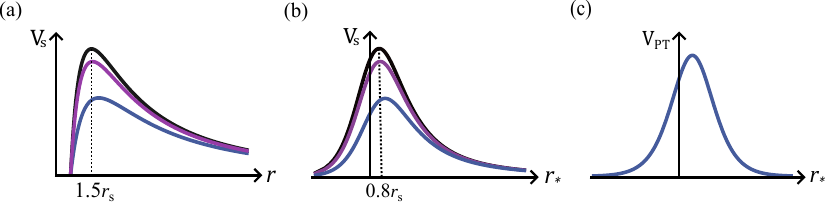}
\caption{Black hole potentials for fields of different spin. (a): radius coordinate, (b): tortoise coordinate; (black: $s=0$, purple: $s=1$, blue: $s=2$), (c): P\"oschl Teller potential.}
\label{fig:bhpotentials}
\end{center}
\end{figure} 
In figure \ref{fig:bhpotentials} (a,b) we see the variation of the potential for different fields. Using the tortoise coordinate creates an almost symmetric potential, similar to the inverted P\"oschl-Teller potential in figure \ref{fig:bhpotentials} (c). 
The decay of the field from the potential is an energy loss mechanism for the black hole, which radiates the field. This process of relaxation is called a black hole ringdown. This phenomenon is not restricted to black holes and can be observed in a variety of oscillatory systems, such as church bells, stringed instruments, damped electrical circuits, polariton superfluids \cite{alpeggiani2017, kristensen2014, yan2018, ching1998, jacquet2023}, supergravity \cite{kumar2021}, Bose-Einstein condensates \cite{martone2018} or surface water gravity waves \cite{patrick2018, torres2020}. An example of a ringdown is presented in figure \ref{fig:bhringdown}. 
A remarkable feature of ringdown is that it is a characteristic of the system, not of the way the system was excited. In figure \ref{fig:bhringdown} there is an initial shock in which the system is perturbed, but the following ringdown phase exposes eigenfrequencies of the system, not the excitation. This is important, as an observation of the ringdown, e.g. of a black hole in a gravitational wave observatory, is a fingerprint that reveals properties of the system like the mass and angular momentum of the black hole. It can be used to test the no-hair theorem \cite{MTWGravitation, gossan2012, isi2019}. Formally, the energy loss is encoded in the system description as an open quantum system with a non-Hermitian Hamiltonian. The Hamiltonian does not describe the system in terms of \textit{normal modes} (real frequencies), but quasinormal modes (QNMs), which carry complex frequencies. Thus, the system has an open boundary with the boundary condition:
\begin{equation}
\psi(r)\sim e^{\pm i \omega r_*} \quad \quad \mathrm{as} \quad \quad r_*\rightarrow \pm \infty.
\label{eq:boundarycondition}
\end{equation}
The set of complex frequencies, the eigenvalues of the Hamiltonian, form the QNM spectrum. Both, normal and quasinormal modes, describe linear systems, but the QNMs do not form a basis for the solution space \cite{Berti2009, Lalanne2018, Torres2018}, as there is no spectral theorem. Instead, they well describe the system at intermediate times during the ringdown phase. At this stage the initial perturbation has been absorbed into the QNMs and they have not yet fully decayed. At late times, many systems also feature an oscillatory tail that decays slower than exponential and therefore is not composed of QNMs. A common feature of these modes is that they decay everywhere in time. However, since the energy is not really lost, but propagating to infinity, this can only be a solution if the wavefunction diverges in space. Thus, strictly speaking we are left with unphysical solutions and the QNMs can only be realized in a finite region of space. The QNM spectrum is a discrete set of modes $\omega_n$ ($n=0,1,2,...$), the lowest of which ($n=0$), the fundamental mode, is the least damped compared to the other `overtones' ($n>0$). In an alternative way, these modes can be seen as the set of poles of the system's Greens function \cite{ching1998}.
The calculation of the QNM spectra normally is a numerical procedure. Exact analytical solutions that are desirable in the analysis are possible for a very limited number of potentials. Approximations are useful if they reproduce the lowest lying QNMs, as these are the ones that are most likely detectable. A prominent technique is the `P\"oschl-Teller approximation' \cite{blome1984, schutz1985, konoplya2011}. This approximation recognizes that the lowest order QNMs are not affected by the asymptotic behavior of the potential ($r_*\rightarrow \pm \infty$), but are dominated by the curvature at the top of the potential. Thus we can approximate the potential with a different potential that has analytical solutions, as long as it has the same curvature. The potential chosen for this approximation is the (inverted) P\"oschl-Teller potential \cite{poschl1933} displayed in figure \ref{fig:bhpotentials} (c). The potential and its QNM frequencies are given by \cite{konoplya2011}:
\begin{eqnarray}
V_{\rm PT}(r_*)&=&V_0\, {\rm sech^2}(\alpha \left(r_*-r_{*0})\right)
\label{eq:poeschltellerpotential}\\
\omega_n &=& \pm \sqrt{V_0-\frac{\alpha^2}{4}}-i \alpha \left( n+\frac{1}{2} \right) \quad \quad n=0,1,2...,
\label{eq:bhqnms}
\end{eqnarray}
where $\alpha>0$ determines the curvature of the potential. We can see that the QNM frequency has a negative imaginary part and thus the wavefunction decays in time for all $n$. With each increase in $n$ the amplitude decay is faster by the factor $e^{-\alpha t}$.  The QNM wavefunctions are available analytically as well, e.g. \cite{burgess2024}. We now have set the foundation to find a laboratory system that may serve as an analogue to the QNMs of black holes.

\subsection{Optical analogy to the light ring}
\label{sec:opticallightring}
At the heart of black hole oscillations is (in most cases) the Schr\"odinger-type equation (\ref{eq:schrodingerequation}) with a potential that characterizes the black hole spacetime. An analogy should thus model the potential or centrifugal barrier outside the black hole, in contrast to the analogy in section \ref{sec:fibereventhorizon}. In fact, this new analogy  also applies to dense gravitating objects that have a light ring, but no event horizon, such as collapsing stars. However, to avoid complexity, we focus on black holes.  The inverted P\"oschl-Teller potential (\ref{eq:poeschltellerpotential}) is of particular importance and has a $\rm sech^2$-shape, reminding us of the soliton solution to the nonlinear Sch\"odinger equation (\ref{soliton}). In fact, in section \ref{ssec:raytracing} we had seen that ray trajectories are indicative of the soliton creating a repulsive potential. However, we need to find out which type of wave equation the solution follows. For this, we go back to the perturbation equation (\ref{eq:propagationequationRR}). We approximate the first operator on the left to $2\beta(i\p_\tau +\omega_0)$, which amounts to neglecting the counterpropagating modes. Also neglecting the fast oscillating and non-phasematched last term, we arrive at: 
\begin{align}
i\p_\zeta {E_p}+\hat{D}{E_p}+2\hat{\gamma}(i \p_\tau+\omega_0)|E_s|^2 E_p  =0.
 \label{eq:qnmperturbationequation1}
\end{align}
This equation describes perturbations around the soliton. For greater generality, we move the reference frequency of this equation to the carrier frequency $\omega_{ p}$ of the perturbation field: 

\begin{align}
E_p(\tau, \zeta)=a(\tau, \zeta) e^{i(\beta(\omega_p)-\beta_0)z+i(\omega_0-\omega_p)t}=a(\tau, \zeta) e^{i[(\beta(\omega_p)-\beta_0)+\beta_{\rm 01}(\omega_0-\omega_p)]\zeta+i(\omega_0-\omega_p)\tau},
 \label{eq:carrierchange}
\end{align}
and inserting this ansatz we arrive at: 
\begin{align}
\p_\zeta {a}-(\beta_{\rm 01}-\beta_{\! p1}) \p_\tau a -i \sum_{n=2}^{\infty} \frac{\beta_{\! pn}}{n!} (i\p_\tau)^n a-2i\hat{\gamma}(i \p_\tau+\omega_{ p})|E_s|^2 a  =0,
 \label{eq:qnmperturbationequation2}
\end{align}
where we also have expanded the dispersion $\hat{D}$ in (\ref{eq:qnmperturbationequation1}) around $\omega_{ p}$. The shift of carrier allows us to neglect dispersion higher than second order for the probe around $\omega_p$ in much the same way as we did for the soliton. We also neglect the dispersion of the nonlinearity ($\gamma(\omega)\rightarrow \gamma_0$), insert the soliton solution, equation (\ref{soliton}), and assume that the perturbation central frequency is of the same group velocity as the soliton to remove the second term. We obtain:
\begin{align}
-i\p_\zeta {a}+  \frac{\beta_{\! p2}}{2} \p_\tau^2a -2 {\gamma_0}P_0 \,{\rm sech}^2(\tau/T_0)\, a  =0,
 \label{eq:solitonschroedinger}
\end{align}
which demonstrates that the perturbation follows a Schr\"odinger-type equation such as the field in the gravitational case (cf. equation (\ref{eq:schrodingerequation})). This is a noteworthy result. In comparison to the Schr\"odinger equation in quantum mechanics $\tau$ represents space and $\zeta$ represents time. Furthermore,  $\beta_{ p2}$ plays the role of the mass, while ${\gamma_0}P_0 \,{\rm sech}^2$ represents the potential and is positive. The second and third terms can have equal or opposite signs, depending on the sign of $\beta_{ p2}$: If it is negative, the equation maps to a Schr\"odinger equation with attractive potential, whereas the potential is repulsive otherwise. As quasinormal modes are defined for repulsive potentials only, perturbations in the anomalous dispersion regime, such as close to the soliton carrier frequency, do not lead to QNMs, but to a stable situation. It can be shown that the perturbed field evolution is still oscillatory in this case, but does not decay exponentially \cite{yang2010}. If the group velocity dispersion of the perturbation is, however, positive (normal dispersion), equation (\ref{eq:solitonschroedinger}) is a time-reversed Schr\"odinger equation with an inverted P\"oschl-Teller potential. 

To compare this to the black hole problem and to find the QNM spectrum, we use a mode ansatz for (\ref{eq:solitonschroedinger}): $a(\tau, \zeta)= u(\tau) e^{-i\Omega_s \zeta}$, where $u(\tau)$ describes the profile of the mode of frequency $\Omega_s$. With the soliton condition (\ref{eq:solitoncondition}) we obtain:
\begin{align}
-\p^2_\tau {u}+\left[  \frac{2 \Omega_s}{\beta_{\!p2}}+4 \frac{|\beta_{02}|}{\beta_{\!p2} T^2_0} \,{\rm sech}^2(\tau/T_0)\,\right] u  =0,
 \label{eq:qnmmodeunscaled}
\end{align}
which writes in the scaled `soliton frame' $\tau'=\tau/T_0, \zeta'=\zeta \,|\beta_{02}|/T\testcmd{2}{0}$ as:
\begin{align}
-\p^2_{\tau'} {u}+ 2 \, \frac{|\beta_{02}|}{\beta_{\!p2}}\left[  \Omega_s+2 \,{\rm sech}^2(\tau')\,\right] u  =0.
 \label{eq:solitonqnmmodeequation}
\end{align}
We can compare this with the black hole mode equation ($\omega =  \Omega_{\rm bh};\, \rho=\alpha(r_*-r_{*0})$) \cite{konoplya2011}:
\begin{align}
-\p^2_{\rho} {\psi}+  \,\left(  -\frac{\Omega^2_{\rm bh}}{\alpha^2}+\frac{V_0}{\alpha^2}\, \rm{sech}^2(\rho)\right) \psi  =0.
 \label{eq:bhmodeequation}
\end{align}
This allows the identifications listed in table (\ref{tab:dictionary}).

\begin{table}
\begin{center}
\begin{tabular}{|>{\arraybackslash}m{3cm}|>{\centering\arraybackslash}m{4.5cm}|>{\centering\arraybackslash}m{4.5cm}|} \hline
& Soliton QNM & Black Hole QNM \\ \hline \hline
time & $-\zeta'$ & $t$ \\ \hline 
space & $\tau'$ & $\rho=\alpha (r_*-r_{*0})$ \\ \hline 
\rule{0pt}{6ex}%
$\rm mode \vphantom{\dfrac{A}{A}}$&
$2 \, \frac{|\beta_{02}|}{\beta_{\!p2}} \Omega_s \vphantom{\dfrac{A}{A}}$ &
$-\dfrac{\Omega^2_{\rm bh}}{\alpha^2} \vphantom{\dfrac{A}{A}}$ \\ \hline
\rule{0pt}{5ex}%
potential &
$4\frac{|\beta_{02}|}{\beta_{\!p2}} \,\rm{sech}^2 \vphantom{\dfrac{A}{A}}$ &
$\frac{V_0}{\alpha^2} \,\, \rm{sech}^2 \vphantom{\dfrac{A}{A}}$ \\ \hline  
\end{tabular}
\caption{Corresponding quantities for soliton- and black hole QNMs.}
\label{tab:dictionary}
\end{center}
\end{table}
By comparison we see that the two equations are formally equivalent. This means that the mode profiles $\psi(\rho)$ and $u(\tau')$ are exactly identical and also that the expansion of arbitrary fields in the mode profiles is identical.
We note, however, that the QNM frequencies are not identical, but instead are related by:
\begin{align}
\Omega_s=-\frac{\beta_{\!p2}}{2\alpha^2 |\beta_{02}|} \,\Omega^2_{\rm bh}.
 \label{eq:frequencytranslation}
\end{align}
The different mode frequencies imply that the fields evolve differently in time, and thus have different temporal profiles. We will now turn our attention to the actual mode solutions for the soliton QNMs and get back to the differing temporal behavior.
\subsection{Soliton quasinormal modes}
\label{sec:solitonqnms}
\begin{figure}
    \centering
    \includegraphics[scale=0.65]{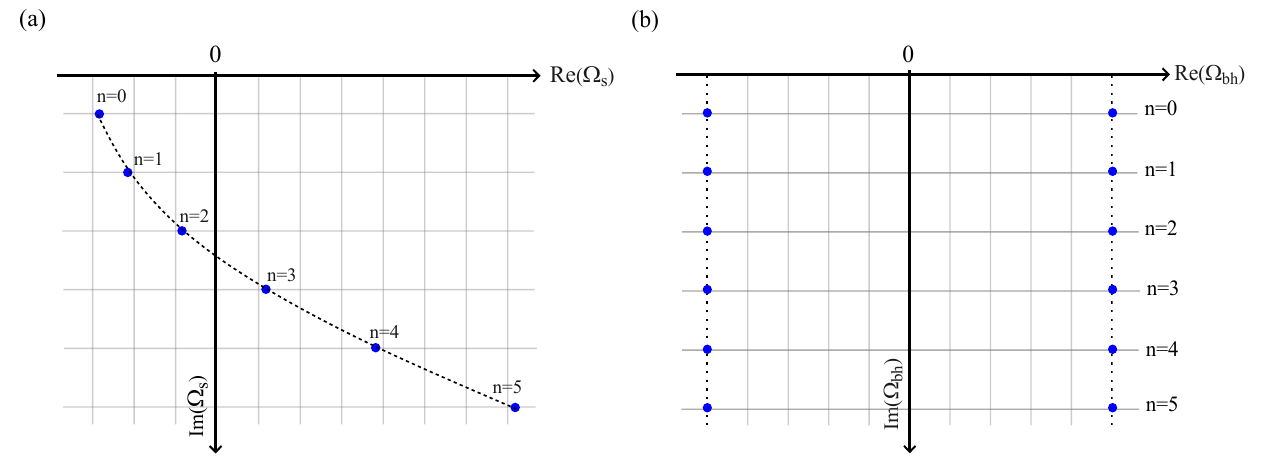}
    \caption{(a) Spectrum of soliton quasinormal modes. $n$ is the overtone index and all modes lie on a parabola in the complex plane, shown here for a group-velocity dispersion ratio $\frac{\beta_{\!p2}}{|\beta_{02}|}=0.45$. (b) Spectrum of black hole quasinormal modes. The spectra transform into each other with relation (\ref{eq:frequencytranslation}). }
    \label{fig:solitonqnmspectrum}
\end{figure}
\begin{figure}[H]
\begin{center}
\includegraphics[scale=0.4]{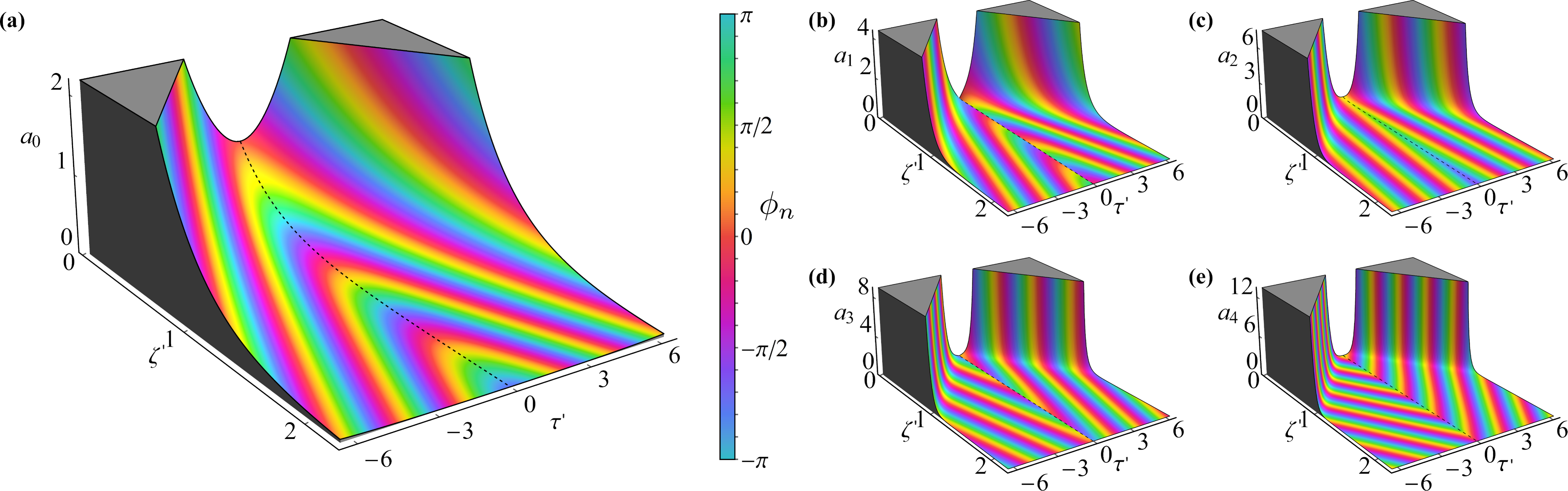}
\caption{The quasinormal modes of solitons. The fundamental mode (a) and the four lowest overtones (b)-(e) ($Q=2.78$). All modes similarly decay in time $\zeta'$ and diverge in space $\tau'$. The phase is indicated by the coloring, which shows that the phase velocity is changing size and direction with overtone index $n$ \cite{burgess2024}.} 
\label{fig:qnms}
\end{center}
\end{figure}

The QNM fields are the solutions of the mode equation (\ref{eq:solitonqnmmodeequation}). The P\"oschl-Teller potential has analytic solutions with the following spectrum \cite{konoplya2011}:
\begin{align}
\Omega_{sn}=\frac{\beta_{\!p2}}{2 |\beta_{02}|} \left[ \left( n+\frac{1}{2} \right)^2 -\left( \frac{4|\beta_{02}|}{\beta_{\!p2} }-\frac{1}{4}\right)\right]-i\frac{\beta_{\!p2}}{ |\beta_{02}|}\left( n+\frac{1}{2} \right)\sqrt{ \frac{4|\beta_{02}|}{\beta_{\!p2} }-\frac{1}{4}}, \quad n=1,2,...
 \label{eq:solitonqnmfrequencies}
\end{align}
Figure \ref{fig:solitonqnmspectrum} (a) shows this spectrum in the complex plane. The discrete spectrum (\ref{eq:solitonqnmfrequencies}) is determined by a single positive parameter, the ratio of group velocity dispersions. We see immediately that $\beta_{\!p2}$ must be positive to generate damped modes, i.e. QNMs. Then, the overtones are increasingly damped with the damping proportional to the overtone index, as seen in the imaginary part, leading to equally spaced modes vertically in figure \ref{fig:solitonqnmspectrum} (a). Figure \ref{fig:solitonqnmspectrum} (b), in comparison, shows the black hole QNM spectrum, which can be retrieved by relation (\ref{eq:frequencytranslation}).
The full QNM mode for the soliton is the solution of the Schr\"odinger equation (\ref{eq:solitonqnmmodeequation}), subject to the boundary conditions. In relativistic systems the boundary conditions are given by outgoing waves, characterized by either phase or group velocity, as these are identical in a non-dispersive system. In the dispersive optical soliton analogue, these boundary conditions relate to the group velocity \cite{burgess2024}, which is also the velocity of energy leaving the system. The explicit modes are \cite{burgess2024}:
\begin{eqnarray}
a_n(\tau', \zeta')=\mathcal{A} \cosh^{n+1/2}(\tau')\, e^{{\rm Im}(\Omega_{sn})\zeta'}  f_n{(\tau')}\, e^{i\phi_n(\tau', \zeta')},
 \label{eq:solitonqnm}
\end{eqnarray}
where $\mathcal{A}$ is an amplitude and
\begin{eqnarray}
f_n(\tau')&=&_{2}F_1\big[-n,2\,i\, Q-n;\frac{1}{2}+i\,Q-n;\left[1-\tanh(\tau')\right]/2],\quad \nonumber \\
\phi_n(\tau', \zeta')&=&-\,Q \ln\big[\cosh(\tau')\big]-{\rm Re}(\Omega_{sn}) \zeta' \quad {\rm and} \nonumber \\
Q&=&\sqrt{\frac{4|\beta_{02}|}{\beta_{\!p2}}-\frac{1}{4}}. \nonumber 
\end{eqnarray}
Here $_{2}F_1$ is the hypergeometric function. The qualitative behavior of the modes, which is also displayed in figure \ref{fig:qnms}, can be understood directly from equation (\ref{eq:solitonqnm}): The first factor describes the divergence in space $\tau'$ and the second factor the decay in time $\zeta'$, both increasing with overtone $n$. $f_n$ is polynomial for integer $n$. The phase $\phi_n$ sets the phase velocity, depending on the sign of ${\rm Re}(\Omega_{sn})$, such that we change from outgoing to ingoing phase velocity between $n=2$ and $n=3$ in this example.   
The change in phase velocity is a departure from the black hole QNM behavior due to dispersion in our optical system. While this is an interesting fact that indicates the modifications of dispersion on the QNM spectra, we can define a regime in which dispersion is negligible for the lowest modes. For this, we take the limit ${|\beta_{02}|}\gg{\beta_{\!p2}}$ in the soliton QNM spectrum (\ref{eq:solitonqnmfrequencies}), which then simplifies to:
\begin{eqnarray}
\Omega_{sn}=-2-i\sqrt{\frac{4\beta_{\!p2}}{|\beta_{02}|}}\left( n+\frac{1}{2}\right),
 \label{eq:simulatorspectrum}
\end{eqnarray}
which is now formally equivalent to the black hole spectrum (\ref{eq:bhqnms}). This means that in this regime the optical analogue is reproducing the black hole case in space and time and can serve as a black hole simulator. For the black hole, this simulator regime is reached if $V_0\gg\alpha^2$, i.e. for black holes in the eikonal regime (see table (\ref{tab:dictionary})).   
\begin{figure}[H]
\begin{center}
\includegraphics[scale=0.7]{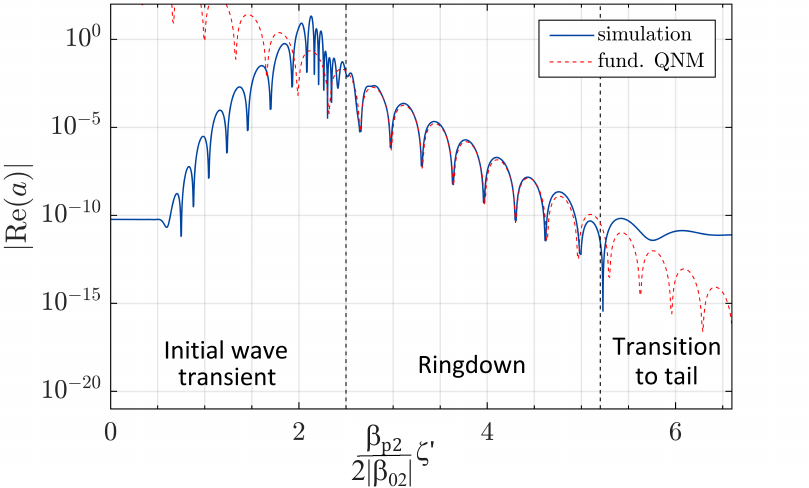}
\caption{The ringdown of an optical soliton: The electric field of the perturbation (of unit amplitude) of the excited soliton is recorded at a fixed point in transmission (numerical simulation with $Q=3.12$) \cite{burgess2024}.}
\label{fig:solitonringdown}
\end{center}
\end{figure} 
Finally, the ringdown of the soliton can be observed if we excite the soliton quasinormal mode oscillations by colliding the soliton with a pulse. This is demonstrated in figure \ref{fig:solitonringdown}, where the perturbative field is simulated at a position $\tau_0$ in the wake of the transmitted part of the pulse \cite{burgess2024}. The general behavior is not critically dependent on any simulation parameters. We can observe that the field is initially perturbed due to the collision, but then starts to decay exponentially with a frequency well characterized by the fundamental QNM (red line). This is the characteristic ringdown of the system. At late time, the decay transitions into a slower decay due to dispersion.

\section{Conclusion}
In these lectures we have reviewed the fundamentals of the most prominent optical analogue systems of nonlinear fiber optics. For the widely used single mode step-index profile fiber, we have analyzed the linear and nonlinear propagation of light and derived a number of variants of the wave propagation equation (\ref{eq:propagationequationtime}), each of which are useful to describe certain wave phenomena in the medium. In a major simplification we have neglected the scattering of light on optical phonons in the fiber, which is called the Raman effect. Except for the shortest pulses below $100\,$fs or for longer propagation distances, this works well. But beyond this work, ultrashort pulses that are influenced by the Raman effect, lead to frequency shifting due to the soliton self-frequency shift \cite{gordon1986,mollenauer1986}, which in turn leads to an acceleration of the pulses, leading to another rich class of physics in this relativistically moving non-inertial frame \cite{gorbach2007,hill2009}. We have then presented a concise treatment of the fiber-optical Hawking effect and closely related phenomena. Finally, the soliton was presented as a system that can simulate the centrifugal barrier of a black hole or otherwise dense object and we have derived the complex frequency spectrum of the quasinormal modes. 
With this stage set, we expect that optical systems will be able to support a number of classical gravity analogues. Their respective quantum signatures are relatively accessible and can serve as guide stars for quantum effects in gravity. This will inspire the discovery of novel fiber phenomena such as negative frequency resonant radiation and new questions about how gravitational systems behave. For the analogue systems presented here, this includes the investigation of spectral instabilites and pseudospectra \cite{Lalanne2018,burgess2024} or the quantum backreaction problem \cite{baak2025}, both of which are difficult to calculate numerically. Beyond these effects, the physics of the quantum vacuum contains many linear and nonlinear phenomena, most of which are undetected to date and can serve for inspiration to find new analogies in optics \cite{Schutzhold2025}. The time is ripe to expand the portfolio of experiments far beyond the Hawking effect.

\section*{Acknowledgments}
We gratefully acknowledge useful discussions with Christopher Burgess and Sangshin Baak. 

\textit{Funding information:}
This work was supported in part (F.K.) by the Science and Technology Facilities Council through the UKRI Quantum Technologies for Fundamental Physics Program [Grant ST/T005866/1 and ST/Y004361/1]. A. Z. was supported by the UK Engineering and Physical Sciences Research Council [Grant No.  EP/W524505/1 ]. D.K. acknowledges financial support by DIM ORIGINES program from {\textit{R\'egion \^Ile de France}} and from the NSF Grant No. PHY-2110273. D.K. also received support through the Atracci\'{o}n de Talento Cesar Nombela grant No 2023-T1/TEC-29023, funded by Comunidad de Madrid (Spain); as well as financial support via the Spanish Grant PID2023-149560NB-C21, funded by MCIU/AEI/10.13039/501100011033/FEDER, UE. For the purpose of open access, the author has applied a Creative Commons Attribution (CC BY) license to any Author Accepted Manuscript version arising.

\begin{appendix}

\section{ Condition for the wave equation analogy}
Here we will show under which conditions the propagation of light in a fiber around the soliton can be described by an equation in the form of the Klein-Gordon equation.
We have seen that the wave equation (\ref{eq:wave_eqn_from_metric}) obtained from the Klein-Gordon equation (\ref{eq:KG.eqn}) is close to the wave equation in optics:
\begin{equation}
c^2\partial_z^2 A_p-\p_\tau \left( n^2_{\rm total} \p_\tau A_p \right) + \left( \p_\tau \ln n_{\rm total} \right) \left[ c^2 \beta_1 \p_z A_p + n^2_{\rm total} \p_\tau A_p \right]=0. \label{eq:Awave_eqn_from_metric}
\end{equation}
The first two terms lead to exactly the same equation as in the optical case. To find out, under which conditions the remaining term is negligible, we use a geometric optics (eikonal) approximation as in section \ref{ssec:raytracing}. This is fitting, as the metric should describe the propagation of rays (null geodesics) correctly. We start with the ansatz 
\begin{equation}
E_p=-\p_\tau A_p = e^{i(\beta(\omega) z - \omega t)}, \label{eq:Aeikonal anstz}
\end{equation}
which leads to 
\begin{equation}
\beta^2(\omega) - 2 \frac{n^2_{\rm total} \omega}{c^2} \left( i\p_\tau \ln n_{\rm total} \right)- \frac{n^2_{\rm total} \omega^2}{c^2}+\left( i\p_\tau \ln n_{\rm total} \right) \left[ \frac{n^2_{\rm total} \omega}{c^2} -\beta_1 \beta(\omega)  \right] =0. \label{eq:Aeikonal_ansztz_in_wave_equation}
\end{equation}
We define a small quantity $\varepsilon = \left( i\p_\tau \ln n_{\rm total} \right)/\omega $ and solve for $\beta(\omega)$:
\begin{equation}
\beta(\omega)=\frac{n_{\rm total} \omega}{c} \left[ \frac{\varepsilon}{2} \frac{\beta_1 c}{ n_{\rm total}}\pm \sqrt{1+\varepsilon +\frac{\varepsilon^2}{4} \left( \frac{\beta_1 c}{n_{\rm total}}\right)^2} \right]. \label{eq:Aeikonal_dispersion_equation}
\end{equation}
Although there are two solutions, we choose the positive-sign solution only as it corresponds to light propagating forward in the fiber. We observe that $ \frac{\beta_1 c}{n_{\rm total}} \approx 1$. Thus, for small $\varepsilon$ we approximate to the first two leading orders:
\begin{equation}
\beta(\omega)=\frac{n_{\rm total} \omega}{c} \left( 1 + \varepsilon \right) + \mathcal{O}(\varepsilon^2). \label{eq:Aeikonal_dispersion_equation2}
\end{equation}
This is the eikonal-approximated dispersion relation. We now compare this to the corresponding dispersion relation of the wave equation (\ref{eq:wave_eqn_from_metric}):
\begin{equation}
\beta(\omega)= \frac{n_{\rm total} \omega}{c} \sqrt{ 1 + 2 \varepsilon } =\frac{n_{\rm total} \omega}{c} \left( 1 + \varepsilon \right) + \mathcal{O}(\varepsilon^2), \label{eq:Aeikonal_dispersion_from_wave_equation}
\end{equation}
which is the same relation. Therefore, if $|\varepsilon|\ll 1$, i.e.
\begin{equation}
 \p_\tau \ln n_{\rm total}  \ll \omega, \label{eq:Aanalogue_condition}
\end{equation}
the wave equation is well approximated up to linear order in $\varepsilon$. In other words, the last term in equation (\ref{eq:Awave_eqn_from_metric}) is of higher than linear order in $\varepsilon$. Physically, the condition means that the refractive index modulation by the soliton is slower than the optical period. This condition is very well met, except for soliton pulses approaching a single cycle. 

\section{Gradient of the divergence of the electric field}
\label{sec:div_E}
In section \ref{sec:linear}, we established that the fiber admits, approximately, linearly polarized solutions with a component given by (\ref{HE11}). In theory, both $E_y$ and $E_z$ solutions are present. However, their amplitudes are much smaller than the $x$-component, and the electric field can be approximated by $\bm{E}(\bm{r},t)\approx E_x(\bm{r},t)\hat{x}$. This approximation, however, introduces a divergence term of the electric field, i.e. $\nabla\cdot \bm{E}\neq 0$. Furthermore, we used this solution, derived in the linear regime, to study the nonlinear propagation of light. In this appendix, we show that the contribution of the divergence is smaller than the Laplacian of the electric field, justifying thus, why it is routinely neglected in the literature \cite{Agra-NL} when deriving the NLSE. 

Let us begin by considering a solitonic (complex) solution to (\ref{wave3}), expressed in Cartesian coordinates
\begin{equation}
\bm{E}(x,y,z,t)=\hat{x}\,E_o\,\text{sech}\left(\frac{t-z/v}{T_0}\right)e^{-\frac{x^2+y^2}{w^2}}e^{i(\beta_0 z- \omega_0 t)} \label{eq:Ex_soliton}
\end{equation}
The latter is coming by joining together the envelope (\ref{soliton}) with the transverse Gaussian profile of the $HE_{11}$ mode (\ref{HE11}), and the rapidly oscillatory part $e^{i(\beta_0 z-\omega_0 t)}$, where $\omega_0$ and $\beta_0$ are the carrier frequency and wavenumber, respectively, and $E_o$ is the overall amplitude of the field. The most general wave equation for the propagation of light in a medium without free charges and currents is (\ref{wave1}). The double curl is expressed as
\begin{equation}
 \nabla \times \nabla \times \bm{E}=\nabla(\nabla\cdot \bm{E})-\nabla^2\bm{E}  \label{eq:Laplacian}  
\end{equation}
We now compare the relative strength of the two terms. In particular, using (\ref{eq:Ex_soliton}), we compute
\begin{align}
\nonumber
\frac{\hat{x}\cdot\nabla^2\bm{E}}{E}=\frac{1}{w^4T_0^2v^2}\Big\{&
T_0^2v^2\left[4(x^2+y^2)-4w^2-w^4\beta_0^2\right]\\
&+w^4\left[\text{tanh}\left(\frac{t-z/v}{T_0}\right)\left(2i\beta_0T_0v+\text{tanh}\left(\frac{t-z/v}{T_0}\right)\right)-\text{sech}^2\left(\frac{t-z/v}{T_0}\right)\right]\Big\}   \label{eq:gradient_of_divergence}
\end{align}
\begin{align}
\frac{\hat{x}\cdot\nabla\left(\nabla\cdot\bm{E}\right)}{E}=\frac{2(2x^2-w^2)}{w^4} 
\end{align}
The Laplacian term scales as $\left|\hat{x}\cdot\nabla^2\bm{E}\right|\sim \beta_0^2E$, while the gradient of the divergence scales as $\left|\hat{x}\cdot\bm{\nabla}(\bm{\nabla}\cdot \bm{E})\right| \sim E/w^2$. To see this more explicitly, we compute (\ref{eq:Laplacian}) and (\ref{eq:gradient_of_divergence}) at the center of the soliton $x=y=0, z=vt$
\begin{align}
\left|\frac{\hat{x}\cdot\nabla^2\bm{E}}{E}\right|&=\left|\frac{4}{w^2}-\beta_0^2-\frac{1}{T_0^2v^2}\right|, \label{eq:Laplacian_center}\\
\left|\frac{\hat{x}\cdot\nabla\left(\nabla\cdot\bm{E}\right)}{E}\right|&=\frac{2}{w^2}. \label{eq:gradient_of_divergence_center}
\end{align}
Typical values for the parameters above are: $\beta_0\sim \frac{2\pi}{\lambda_o}\approx 6\times10^6 \text{m}^{-1}$, for a mode with $\lambda_o=1\mu\text{m}$, $u\sim c$, $T\sim 100\, \text{f}s$. The mode radius $w$ is of the order of magnitude of the core radius,  $a\sim 5 \mu\text{m}$. Thus, the dominant term in (\ref{eq:Laplacian_center}) is $\beta_0^2\sim \mathcal{O}(10^{13})\,m^{-2}$, while (\ref{eq:gradient_of_divergence_center}) scales as $\mathcal{O}(10^{11})\, m^{-2}$. In conclusion, one can safely neglect the gradient of the divergence of the electric field as long as $\beta_0\gg w^{-1}$.

\end{appendix}

\bibliography{Refs3}

\end{document}